\documentclass[preprint,preprintnumbers,showkeys,amsmath,amssymb,prb]{revtex4}

\usepackage{graphicx}        
\usepackage{bm}

\newcommand{\erf}{\mbox{erf}}
\newcommand{\erfc}{\mbox{erfc}}
\newcommand{\D}{\mbox{d}}
\newcommand{\I}{\mbox{i}}
\renewcommand{\vec}[1]{\mathbf{#1}}

\usepackage{array}
\usepackage{amsmath}
\begin{document}
\tableofcontents
\newpage

\title{First-principles calculation method and its applications for two-dimensional materials}
\author{Yoshiyuki Egami}
\affiliation{Faculty of Engineering, Hokkaido University, Sapporo, Hokkaido 060-8628, Japan}
\author{Shigeru Tsukamoto}
\affiliation{Peter Gr{\"u}nberg Institut/Institute for Advanced Simulation, Forschungszentrum J{\"u}lich, D-52428 J{\"u}lich, Germany}
\author{Tomoya Ono}
\affiliation{Center for Computational Sciences, University of Tsukuba, Tsukuba, Ibaraki 305-8577, Japan}
%
%
\begin{abstract}
We present details of our effective computational methods based on the real-space finite-difference formalism to elucidate electronic and magnetic properties of the two-dimensional (2D) materials within the framework of the density functional theory. The real-space finite-difference formalism enables us to treat truly 2D computational models by imposing individual boundary condition on each direction. The formulae for practical computations under the boundary conditions specific to the 2D materials are derived and the electronic band structures of 2D materials are demonstrated using the proposed method.
Additionally, we introduce other first-principles works on the MoS$_2$ monolayer focusing on the modulation of electronic and magnetic properties originating from lattice defects.
\end{abstract}

\maketitle

\section{Introduction}
Atomically thick two-dimensional (2D) materials have been increasingly attracting interest from the perspective of fundamental science and advanced engineering because their electronic, magnetic, optical, and chemical properties are significantly different from those of 3D bulk materials. The most widely studied 2D material since the successful exfoliation using the micromechanical cleavage method in 2004 is graphene,\cite{KSNovoselov2004} because of its rich physics and high electron mobility.
However, pristine graphene is a gapless semiconductor, and thus, cannot obtain an effective current on/off ratio, which hinders its application to semiconductor fields such as field-effect transistors (FETs). Although various techniques have been investigated to expand the band gap of bilayer graphene systems\cite{EMcCannPRL2006,EMcCannPRB2006,TOhta2006,EVCastro2007} for developing graphene-based FETs, it is still challenging to obtain the band gap large enough for the industrial applications. Recently, graphene-like 2D materials, e.g., silicene, germanene, and phosphorene, and transition metal dichalcogenide (TMD) monolayer have also attracted much interest due to their easily tunable or intrinsic band gaps and potential applications in next-generation electronics, spintronics and optical devices.\cite{WYLiang1986,AAruchamy1992,ThBoeker2001,AKlein2001,ApplPhysLett98-223107,NanoLett12-113,ACSNano8-4033} However, in order to apply them in the development of new devices, there remain many open questions to be answered regarding their electronic and magnetic characteristics. To resolve the questions, theoretical approaches using first-principles calculations are indispensable as well as experimental investigations.

Conventional first-principles calculation methods are based on basis-set expansion techniques using such as atomic orbitals or plane waves. In these methods, however, we should always pay attention to whether the basis function used satisfies the required calculation accuracy and to the fact that the boundary condition does not correspond to that of the actual experiments. On the other hand, the real-space finite-difference method\cite{PRL72-001240,PRB50-011355,PRB50-012234,PRB53-012071} enable us to avoid the above problems, since the wave function and potential on real-space grids are directly calculated without using basis functions. For instance, the calculation accuracy can be simply improved by narrowing the grid spacing. Moreover, by imposing individual boundary condition on each direction, i.e., a periodic boundary condition in the film direction of the 2D materials and an isolated boundary condition in the direction perpendicular to the film plane, one can treat more strictly 2D computational models.

In this review paper, we present details of numerical procedures based on the real-space finite-difference formalism within the framework of the density functional theory\cite{PhysRev136-00B864} to estimate electronic and magnetic properties of 2D materials. 
The formulae for practical computations under the boundary conditions specific to the 2D materials are derived and the electronic band structures of 2D materials are demonstrated using the proposed method. First, we present a derivation of the discretized Kohn--Sham equation based on the real-space finite-difference formalism. Then, using multipole expansion and Ewald summation techniques,\cite{PRL72-001240,PRB50-011355,ono-ewald} efficient procedures to compute the Hartree potential for 2D materials by solving the Poisson equation are described. Furthermore, we derive the Kohn--Sham equation and the Poisson equation in the Laue representation\cite{PRB59-015609,PRB56-012874} which is well-known as an effective and suitable technique to treat 2D periodic materials and expresses physical quantities using the two-dimensional plane wave expansion and a one-dimensional real-space grid.
Additionally, we demonstrate the band structure calculations for graphene-like 2D materials (graphene, silicene, and germanene) using the proposed method and introduce other first-principles works on electronic and magnetic properties of a MoS$_2$ monolayer.

In the following five sections, our computational formalism is described in detail. In Sect.~\ref{sec:ono-3D Periodic Boundary Condition: Crystals}, the electronic band structure calculations for the several 2D materials are demonstrated by means of our formalism. Section~\ref{sec:egami1} introduces other first-principles studies on the MoS$_2$ monolayer. The summary and outlook for the theoretical study on the 2D materials is presented in Sect.~\ref{sec:conclusion}.

\section{Density functional theory and Kohn-Sham equation}
\label{tsuka-sec:DFTKohnShamEquation}
We would better start this article from brief introduction of two essential background theorems on first-principles calculations, on which the theoretical approach to be stated in this article is based. 
One is the density functional theory, which has been built up by Hohenberg and Kohn in 1964.\cite{PhysRev136-00B864} 
The density functional theory states that the charge density distribution at the ground state of a system minimizes the total energy functional under the constraint of particle number conservation, and determines all the properties of the system, such as eigenvalues and eigenfunctions. 
The total energy functional is, however, in an universal form, and difficult to be evaluated in numerical calculations. 
Therefore, the density functional theory had to include promising approximations of the universal functional, which are adoptable in practical computations. 

In 1965, Kohn and Sham has introduced an approximation with \emph{non-interacting} electrons, which enables us to reproduce the ground-state charge density distributions of complicated many-electron systems only by solving inexpensive single-particle Schr{\"o}dinger-like equation, namely Kohn-Sham equation.\cite{PhysRev140-0A1133}
\begin{equation}
-\frac{1}{2}\nabla^{2}\psi(\vec{r})+v_{\mathrm{eff}}(\vec{r})\psi(\vec{r})=\varepsilon\psi(\vec{r}), 
\label{tsuka-eq:KohnShamEquation}
\end{equation}
where $v_{\mathrm{eff}}(\vec{r})$ represents the Kohn-Sham effective potential and reads
\begin{equation}
v_{\mathrm{eff}}(\vec{r})=\sum_{s}v_{\mathrm{ion}}^{s}(\vec{r})+v_{\mathrm{f}}(\vec{r})+v_{\mathrm{H}}(\vec{r})+v_{\mathrm{xc}}(\vec{r}). 
\end{equation}
Here, the first, second, third, and forth terms in the right-hand side denote ionic core potential, external potential such as electric field, Hartree potential, and exchange-correlation potential, respectively. 
To determine the electronic structure at the ground state, the Kohn-Sham equation (\ref{tsuka-eq:KohnShamEquation}) is, in general, solved as an eigenvalue problem for a certain number of eigenpairs, i.e., eigenenergy $\varepsilon_{i}$ and wave function $\psi_{i}(\vec{r})$ for $i=1,2,\cdots$.
According to the density functional theory, the Hartree potential and the exchange-correlation potential depend on the electron density,
\begin{equation}
\rho(\vec{r})=\sum_{i}n_{i}|\psi_{i}(\vec{r})|^{2}, 
\end{equation}
where $n_{i}$ denotes the occupation number of the $i$th Kohn-Sham orbital. 
The electron density $\rho(\vec{r})$ obviously depends on the wave functions $\psi_{i}(\vec{r})$, which are determined by solving the Kohn-Sham equation (\ref{tsuka-eq:KohnShamEquation}). 
Because of the dependency on each other, this series of equations needs to be solved in a self-consistent manner.

We notice that throughout this article we use the Hartree atomic unit, i.e., $|e|=m=h/2\pi=1$, where $e$, $m$, and $h$ are the electron charge, electron mass, and Planck's constant, respectively. 

\section{Real-space finite-difference formalism}
The real-space finite-difference formalism, to be stated in this section, is one of the methods to solve the Kohn-Sham equation (\ref{tsuka-eq:KohnShamEquation}) within the framework of the density functional theory, and has been at first proposed by Chelikowsky et al. in 1994.\cite{PRL72-001240,PRB50-011355,PRB50-012234,PRB53-012071}
As shown in Fig.~\ref{tsuka-fig:Figure1}(a), the real-space finite-difference formalism represents the three-dimensional (3D) continuous real space as a 3D discrete space filled with equidistant grid points, i.e., each direction in the 3D real space is sampled with a constant grid spacing $h_{i}$ ($i=x$, $y$, and $z$). 
Therefore, physical quantities being continuous in real space, such as effective potential, electron wave function, and electron density distribution, are also discretized, and the values are defined only on the discretized grid points as shown in Fig.~\ref{tsuka-fig:Figure1}(b).
Consequently, we can directly treat the physical quantities by solving the Kohn-Sham equation (\ref{tsuka-eq:KohnShamEquation}) for the values on the grid points.
This is contrastive to the conventional methods using basis function sets such as atomic orbitals or plane waves, which expand the physical quantities using the basis functions and solve the problems for the expansion coefficients.

In this section, we show how to transcribe the Kohn-Sham equation (\ref{tsuka-eq:KohnShamEquation}) into the real-space finite-difference formalism so that one can solve the differential equation in practical numerical computation. 
In the conventional methods using basis function sets, the second derivative with respect to three real-space directions, as seen in the left-hand side of (\ref{tsuka-eq:KohnShamEquation}), can be managed by differentiating the basis functions. 
In contrast, the real-space finite-difference formalism does not adopt any basis function sets, and thus, the second derivative is approximated by finite-difference formulae. 

\subsection{Finite-difference approximation}
In the real-space finite-difference formalism, the wave function $\psi(\vec{r})$ in (\ref{tsuka-eq:KohnShamEquation}) is discretized and the set of the values on the grid points are treated as a vector. 
Accordingly, the operators at the left-hand side in (\ref{tsuka-eq:KohnShamEquation}) acting on the wave function $\psi(\vec{r})$ need to be discretized and defined in a matrix form. 
The effective potential $v_{\mathrm{eff}}(\vec{r})$ is now assumed to be a local operator, and therefore, is simply expressed as a diagonal matrix. 
On the other hand, the kinetic energy operator in the form of the Laplacian $\nabla^{2}$, i.e., the second derivative with respect to the three real-space directions, is approximately represented by a semi-local matrix form. 
This is called finite-difference approximation.  

The representation of the second derivative operator in a matrix form is derived by using the Taylor expansion of a continuous function $f(x)$. 
Let us consider the Taylor expansion of the one-dimensional (1D) function $f(x)$ with respect to a grid point $x_{i}$, and express the function values at the neighboring grid points $x_{i\pm1}=x_{i}\pm h_{x}$ up to the second order of the Taylor expansion. 
\begin{eqnarray}
f_{i-1} & = & f(x_{i-1}) = f(x_{i})-h_{x}f'(x_{i})+\frac{1}{2}h_{x}^{2}f''(x_{i})+{\cal O}(h_{x}^{3}) \\
f_{i}   & = & f(x_{i})   \\
f_{i+1} & = & f(x_{i+1}) = f(x_{i})+h_{x}f'(x_{i})+\frac{1}{2}h_{x}^{2}f''(x_{i})+{\cal O}(h_{x}^{3})
\end{eqnarray}
This set of the equations can be rewritten in the matrix form
\begin{equation}
\left[
\begin{array}{c}
 f_{i-1} \\
 f_{i}   \\
 f_{i+1}
\end{array}
\right]=\left[
\begin{array}{ccc}
 1 & -1 & \frac{1}{2} \\
 1 &  0 & 0           \\
 1 &  1 & \frac{1}{2}
\end{array}
\right]\left[
\begin{array}{c}
f(x_{i}) \\
h_{x}f'(x_{i})   \\
h_{x}^{2}f''(x_{i})
\end{array}
\right].
\label{tsuka-eq:TaylorCoefficientMatrix}
\end{equation}
Operating the inverse of the $3\times3$ matrix in the right-hand side on the equation above from the left and exchanging the sides, one can have the approximate expressions of the zeroth, first, and second derivatives of the function $f(x)$ at the grid point $x_{i}$ as the weighted summations of the function values $f_{i-1}$, $f_{i}$, and $f_{i+1}$. 
\begin{equation}
\left[
\begin{array}{c}
f(x_{i}) \\
h_{x}f'(x_{i})   \\
h_{x}^{2}f''(x_{i})
\end{array}
\right]=\left[
\begin{array}{ccc}
 0            &  1 & 0            \\
 -\frac{1}{2} &  0 & \frac{1}{2} \\
 1            & -2 & 1
\end{array}
\right]\left[
\begin{array}{c}
 f_{i-1} \\
 f_{i}   \\
 f_{i+1}
\end{array}
\right]
\label{tsuka-eq:InverseTaylorCoefficientMatrix}
\end{equation}
The third row just shows the finite-difference approximation of the second derivative using up to the first nearest function values, i.e., approximation order of $N_{\mathrm{f}}=1$. 
\begin{equation}
f''(x_{i})=\frac{1}{h_{x}^{2}}\left(f_{i-1}-2f_{i}+f_{i+1}\right)
\end{equation}
Thus, the quotients to the respective function values for the approximation order $N_{\mathrm{f}}=1$ are found to be $1$, $-2$, $1$ for $f_{i-1}$, $f_{i}$, and $f_{i+1}$, respectively. 
This scheme can be expanded to higher-order derivatives and more neighboring grid points. 
For the generalized case that considers the Taylor expansion up to $2N_{\mathrm{f}}$th order and $N_{\mathrm{f}}$ neighboring grid points at each side of the grid point $x_{i}$, the function values $f(x_{i+j})$ for $j=-N_{\mathrm{f}},\cdots,N_{\mathrm{f}}$ are expressed as
\begin{equation}
f(x_{i+j})=f_{i+j}=\sum_{k=0}^{2N_{\mathrm{f}}}\frac{(jh_{x})^{k}}{k!}\left.\frac{\D^{k}f(x)}{\D x^{k}}\right|_{x=x_{i}}+{\cal O}(h_{x}^{2N_{\mathrm{f}}+1}).
\label{tsuka-eq:HigherOrderTaylorCoefficientMatrix}
\end{equation}
Thus, the $jk$ element of the Taylor coefficient matrix as in (\ref{tsuka-eq:TaylorCoefficientMatrix}) is $t_{jk}=j^{k}/k!$. 
By taking the procedure for changing the equation as from (\ref{tsuka-eq:TaylorCoefficientMatrix}) to (\ref{tsuka-eq:InverseTaylorCoefficientMatrix}), the $m$th derivative of the function $f(x)$ at the grid point $x_{i}$ is expressed as
\begin{equation}
f^{(m)}(x_{i})=\frac{1}{h_{x}^{m}}\sum_{l=-N_{\mathrm{f}}}^{+N_{\mathrm{f}}}\left\{[t_{jk}]^{-1}\right\}_{ml}f_{i+l}.
\end{equation}

Table \ref{tsuka-tbl:Table1} exhibits the real-space finite-difference coefficients of the second derivative for the approximation orders of $N_{\mathrm{f}}=1,\cdots,8$.
Using the sets of the coefficients $c_{l}$, the second derivative of the function $f(x)$ at the $i$th grid point $x_{i}$ is expressed as
\begin{equation}
\left.\frac{\D^{2}}{\D x^{2}}f(x)\right|_{x=x_{i}}=\frac{1}{h_{x}^{2}}\sum_{l=-N_{\mathrm{f}}}^{+N_{\mathrm{f}}}c_{l}f_{i+l}.
\label{tsuka-eq:SecondDerivativeFiniteDifferenceApproximation}
\end{equation}

We notice that since the finite-difference approximation of the second derivative relates the grid points $x_{i}$ and $x_{j}$ only for all $|i-j|\leq N_{\mathrm{f}}$, the Laplacian operator in the matrix form is not local any more, but the non-zero elements appears only within a diagonal band of the matrix. 
Consequently, the Laplacian operator matrix is sparse in the real-space finite-difference representation. 
The sparseness is advantageous to perform matrix--vector multiplications in numerical computation.

At the end of this subsection, we mention the accuracy of the finite-difference approximation. 
Figure \ref{tsuka-fig:Figure2} draws the energy dispersion relations of a plane wave $\exp(\I k_{x}x)$, which are evaluated by using the finite-difference formula $(\ref{tsuka-eq:SecondDerivativeFiniteDifferenceApproximation})$ with the approximation orders of $N_{\mathrm{f}}=1,\cdots,8$ and at a constant grid spacing $h_{x}=1\;a_{\mathrm{B}}$ ($a_{\mathrm{B}}$ is the unit of length in the Hartree atomic unit, i.e., $1\;a_{\mathrm{B}}=0.529\;\mathrm{\AA}$). 
These energy dispersion relations are compared to that obtained by the analytic solution $\frac{1}{2}k_{x}^{2}$.  
It is clearly seen that the finite-difference approximation deviates more from the analytical solution for lower order and for higher wave number. 
Although increasing the approximation order $N_{\mathrm{f}}$ of the finite-difference formula (\ref{tsuka-eq:SecondDerivativeFiniteDifferenceApproximation}) is the most effective way to improve the accuracy, this simultaneously deteriorates the sparseness of the kinetic energy operator matrix, and thus, leads increase in computational cost. 
For a while, we treat only the case of the central finite-difference approximation, i.e., the approximation order of $N_{\mathrm{f}}=1$, for simplicity.

\subsection{Real-space representation of Kohn-Sham equation}
In the preceding subsection, we have discretized the second derivative operator as well as physical quantities, i.e., wave function $\psi(x)$ and effective potential $v_{\mathrm{eff}}(x)$, and thus, the 1D Kohn-Sham equation is ready to be transcribed into a matrix equation form. 
When a real-space calculation domain is uniformly divided into $N$ grid points and an \emph{isolated} boundary condition ($\psi_{i}=0$ for $i<1$ or $N<i$) is imposed, the product of the Kohn-Sham Hamiltonian $-\frac{1}{2}\frac{\D^{2}}{\D x^{2}}+v_{\mathrm{eff}}(x)$ and the wave function $\psi(x)$, like as the left-hand side of (\ref{tsuka-eq:KohnShamEquation}), is written as the matrix--vector product 
\begin{equation}
\left[-\frac{1}{2}\frac{\D^{2}}{\D x^{2}}+v_{\mathrm{eff}}(x)\right]\psi(x)\approx\left[
\begin{array}{ccccc}
\alpha_{1} & \beta      & 0      & \cdots & 0      \\
\beta      & \alpha_{2} &        & \ddots & \vdots \\
0          &            & \ddots &        & 0      \\
\vdots     & \ddots     &        & \ddots & \beta  \\
0          & \cdots     & 0      & \beta  & \alpha_{N} 
\end{array}
\right]\left[
\begin{array}{c}
\psi_{1} \\
\psi_{2} \\
\vdots   \\
\vdots   \\
\psi_{N}
\end{array}
\right],
\label{tsuka-eq:KohnShamLeftHandSideIsolated}
\end{equation}
where $\alpha_{i}=\frac{c_{0}}{2h_{x}^{2}}+v_{\mathrm{eff}}(x_{i})$ and $\beta=\frac{c_{1}}{2h_{x}^{2}}$. 
Here, one can confirm that the operator matrix is band diagonal. 
In the case of imposing a \emph{periodic} boundary condition, wave functions have finite values even outside the calculation domain, i.e., supercell, and thus, the nonlocal kinetic energy operator matrix refers to the wave function values in the neighboring supercells ($i<1$ and $N<i$). 
According to the Bloch's theorem, the wave function values in the neighboring supercells are determined using the Bloch condition $\psi_{i\pm N}=\exp(\pm\I k_{x}L_{x})\psi_i$, where $k_{x}$ and $L_{x}$ are the wave number and supercell length in the $x$ direction, respectively. 
Therefore, the product of the Kohn-Sham Hamiltonian $-\frac{1}{2}\frac{\D^{2}}{\D x^{2}}+v_{\mathrm{eff}}(x)$ and the wave function $\psi(x)$, like as the left-hand side of (\ref{tsuka-eq:KohnShamEquation}) is written as the matrix--vector product
\begin{equation}
\left[-\frac{1}{2}\frac{\D^{2}}{\D x^{2}}+v_{\mathrm{eff}}(x)\right]\psi(x)\approx\left[
\begin{array}{ccccc}
\alpha_{1}  & \beta      & 0      & \cdots & \beta^{(-)} \\
\beta       & \alpha_{2} &        & \ddots & \vdots      \\
0           &            & \ddots &        & 0           \\
\vdots      & \ddots     &        & \ddots & \beta       \\
\beta^{(+)} & \cdots     & 0      & \beta  & \alpha_{N} 
\end{array}
\right]\left[
\begin{array}{c}
\psi_{1} \\
\psi_{2} \\
\vdots   \\
\vdots   \\
\psi_{N}
\end{array}
\right],
\label{tsuka-eq:KohnShamLeftHandSidePeriodic}
\end{equation}
where $\beta^{(\pm)}=\exp(\pm\I k_{x}L_{x})\beta$, and $\exp(\pm\I k_{x}L_{x})$ is called phase factor. 

So far, we have stated the real-space finite-difference representation of Kohn-Sham equation (\ref{tsuka-eq:KohnShamEquation}) based on the 1D central finite-difference formula. 
This argument is able to be extended straightforwardly to the case of 3D Kohn-Sham equation with higher order of the finite-difference approximation. 
Substituting the finite-difference formula (\ref{tsuka-eq:SecondDerivativeFiniteDifferenceApproximation}) into the Kohn-Sham equation (\ref{tsuka-eq:KohnShamEquation}), the discretized Kohn-Sham matrix equation in the real-space representation consequently reads
\begin{equation}
-\frac{1}{2}\sum_{l=-N_{\mathrm{f}}}^{+N_{\mathrm{f}}}\left[\frac{c_{l}}{h_{x}^{2}}\psi_{i+l,j,k}+\frac{c_{l}}{h_{y}^{2}}\psi_{i,j+l,k}+\frac{c_{l}}{h_{z}^{2}}\psi_{i,j,k+l}\right]+v_{\mathrm{eff},i,j,k}\psi_{i,j,k}=\varepsilon\psi_{i,j,k}.
\label{tsuka-eq:FiniteDifferenceApproximation}
\end{equation}
The subscript $i$, $j$, and $k$ of the wave function and effective potential stand for the real-space grid indexes in the $x$, $y$, and $z$ directions, respectively.

Throughout this section, we have treated the Kohn-sham equation under the assumption of \emph{local} effective potential. 
Even in the case of \emph{nonlocal} effective potential, which is introduced by adopting sophisticated pseudopotential methods,\cite{PRB26-004199,PRB43-001993,PRB41-007892,PRB50-017953} the discretized Kohn-Sham matrix equation including nonlocal effective potential can be derived within the framework of the real-space finite-difference formalism in the similar manner.\cite{PRB72-085115,KikujiHirose2005}

\section{Hartree potential for 2D periodic boundary condition}
\label{sec:fuzzy cell decomposition and multipole expansion technique}
Considering two-dimensional (2D) materials, such as graphene, MoS$_{2}$ monolayer, and Bi bilayer, one can easily see that the 2D materials have periodicity only in the directions of their geometrical extension, which are hereafter referred to as $x$ and $y$ directions. In the $z$ direction perpendicular to the film plane, no periodicity exists and rather they are isolated, as illustrated in Fig.~\ref{tsuka-fig:Figure3}.
The Hartree potential in the Kohn-Sham equation (\ref{tsuka-eq:KohnShamEquation}), which is the Coulomb interaction of electrons, should be carefully treated under the periodic boundary condition because the integration of Coulomb interactions diverges if the integrations are executed infinitely. The Hartree potential is commonly evaluated by solving the Poisson equation.
\begin{equation}
\nabla^{2}v_{\mathrm{H}}(\vec{r})=-4\pi\rho(\vec{r})
\label{tsuka-eq:VHPoissonEquation}
\end{equation}
Equation~(\ref{tsuka-eq:VHPoissonEquation}) is usually solved using conjugate-gradient or steepest-decent method. In the case of the systems including the isolated boundary condition, the boundary values of Hartree potential just outside of the calculation domain are required to solve the Poisson equation. In this section, we introduce an efficient procedure to determine the boundary values of Hartree potential.

The numerical summation over the grids is the most direct procedure to compute the Hartree potential. However, the numerical summation is time-consuming for large systems because the computational cost for the direct summation is proportioned to the 5/3 power of system size. The procedure using a multipole expansion of the electron density around an arbitrary point $\vec{r}^0$ is proposed by Chelikowsky et al.\cite{PRL72-001240,PRB50-011355}
\begin{eqnarray}
v_{\mathrm{H}}(\vec{r})&=& \int \frac{\rho(\vec{r}')}{|\vec{r}-\vec{r}'|} d\vec{r}' \nonumber \\
&=&\sum_{l=0}^\infty \int \frac{\rho(\vec{r}')}{|\vec{r}-\vec{r}^0|} \left(\frac{|\vec{r}'-\vec{r}^0|}{|\vec{r}-\vec{r}^0|}\right)^lP_l(\cos \theta') d \vec{r}' \nonumber \\
&=&\frac{\int \rho(\vec{r}') d \vec{r}'}{|\vec{r}-\vec{r}^0|}+\sum_{\mu} p_\mu \cdot \frac{(r_\mu-r_\mu^0)}{|\vec{r}-\vec{r}^0|^3} \nonumber \\
&& + \sum_{\mu,\nu} q_{\mu \nu} \cdot \frac{3\:(r_\mu-r_\mu^0)(r_\nu-r_\nu^0)-\delta_{\mu \nu}|\vec{r}-\vec{r}^0|^2}{|\vec{r}-\vec{r}^0|^5} \nonumber \\
&& + \cdots
\label{eqn:poisson01}
\end{eqnarray}
with $\mu$ and $\nu$ being $x$, $y$ and $z$. Here, the functions $P_l(\cos \theta')$ $(l=0, 1, 2, \cdots)$ are the Legendre polynomials and $\cos \theta'$ is described as
\begin{equation}
\cos \theta'=\frac{(\vec{r}-\vec{r}^0) \cdot (\vec{r}'-\vec{r}^0)}{|\vec{r}-\vec{r}^0| \cdot |\vec{r}'-\vec{r}^0|}.
\end{equation}
In addition, $p_\mu$ and $q_{\mu \nu}$ are the dipole moment
\begin{equation}
p_\mu=\int (r_\mu'-r_\mu^0) \rho(\vec{r}') d \vec{r}'
\end{equation}
and quadrupole moment
\begin{equation}
q_{\mu \nu}=\int \frac{1}{2} (r_\mu'-r_\mu^0)(r_\nu'-r_\nu^0) \rho(\vec{r}') d \vec{r}' ,
\end{equation}
respectively. Although the computational cost is proportioned to system size in this scheme, the accuracy of the solution largely depends on the choice of the position $\vec{r}^0$.

To avoid the problem in accuracy, the positions of atoms are chosen as the centers of the multipole expansion. This method is called the fuzzy cell decomposition and multipole expansion.\cite{PRB72-085115} A weighting function $\omega_s(\vec{r})$ for the multiple-center system centered at the $s$th nucleus is introduced to decompose the electron density.
\begin{equation}
\rho_s(\vec{r})=\rho(\vec{r})\omega_s(\vec{r}),
\end{equation}
where
\begin{equation}
\rho(\vec{r})=\sum_s\rho_s(\vec{r}).
\end{equation}
$\omega_s(\vec{r})$ is the defining function of so-called Voronoi polyhedra $\Omega_s$,\cite{ono-voronoi} which provides Wigner-Seitz cells and is set to satisfy the following equations.
\begin{equation}
\label{eqn:ono-poisson08}
\sum_s \omega_s(\vec{r}) = 1,
\end{equation}
where
\begin{equation}
\label{eqn:ono-poisson09}
\omega_s(\vec{r}) = \left \{
\begin{array}{ll}
1 & \hspace {5mm} \in \Omega_s \\
0 & \hspace {5mm} \mbox{otherwise}
\end{array}
\right. .
\end{equation}
The multipole expansion for each $\rho_s(\vec{r})$ centered around the position of each nucleus $\vec{R}^s$ gives the Hartree potential 
\begin{eqnarray}
\label{eqn:ono-poisson10}
v_{\mathrm{H}}(\vec{r}) &=& \sum_s \Biggl( \frac{\int \rho_s(\vec{r}') d \vec{r}'}{|\vec{r}-\vec{R}^s|} + \sum_{\mu} p_\mu^s \cdot \frac{r_\mu-R^s_\mu}{|\vec{r}-\vec{R}^s|^3} \nonumber \\
&&+ \sum_{\mu,\nu} q_{\mu \nu}^s \cdot \frac{3(r_\mu-R^s_\mu)(r_\nu-R^s_\nu)- \delta_{\mu \nu}|\vec{r}-\vec{R}^s|^2}{|\vec{r}-\vec{R}^s|^5} + \cdots \Biggr) ,
\end{eqnarray}
with $\mu$ and $\nu$ being $x$, $y$ and $z$. Here, $p_\mu^s$ and $q_{\mu \nu}^s$ are
\begin{equation}
p_\mu^s=\int (r_\mu'-R_\mu^s) \rho_s(\vec{r}') d \vec{r}'
\end{equation}
and
\begin{equation}
q_{\mu \nu}^s=\int \frac{1}{2} (r_\mu'-R_\mu^s)(r_\nu'-R_\nu^s) \rho_s(\vec{r}') d \vec{r}' .
\end{equation}

If $\omega_s(\vec{r})$ is in the manner of a step function at the boundary of $\Omega_s$, the expansion of (\ref{eqn:ono-poisson10}) requires many terms. For rapid convergence of the expansion with respect to the number of the terms in (\ref{eqn:ono-poisson10}), the behavior of $\omega_s(\vec{r})$ near the boundary should be made as smooth as possible using the fuzzy cell technique,\cite{ono-fuzzy} in which the section of the boundary of (\ref{eqn:ono-poisson09}) is fuzzy. The multipole expansion up to the quadrupole is sufficient to obtain an accurate solution when the fuzzy cell is employed.

For later convenience, we define the following quantities.
\begin{equation}
V^{1,s}_{\mathrm{H}}(\vec{r})=\frac{1}{|\vec{r}-\vec{R}^s|},
\end{equation}
\begin{equation}
V^{2,s}_{{\mathrm{H}},\mu}(\vec{r})=\frac{r_\mu-R^s_\mu}{|\vec{r}-\vec{R}^s|^3},
\end{equation}
and
\begin{equation}
V^{3,s}_{{\mathrm{H}},\mu \nu}(\vec{r})=\frac{3(r_\mu-R^s_\mu)(r_\nu-R^s_\nu)- \delta_{\mu \nu}|\vec{r}-\vec{R}^s|^2}{|\vec{r}-\vec{R}^s|^5}.
\end{equation}
Thus, (\ref{eqn:ono-poisson10}) is written as
\begin{eqnarray}
\label{eqn:ono-ewald2D06}
v_{\mathrm{H}}(\vec{r}) &=& \sum_s \left( \int \rho_s(\vec{r}') d \vec{r}' \cdot V^{1,s}_{\mathrm{H}}(\vec{r}) + \sum_{\mu=x,y,z} p_{s,\mu} \cdot V^{2,s}_{{\mathrm{H}},\mu}(\vec{r}) \right. \nonumber \\
&& \hspace{2cm} \left. + \sum_{\mu,\nu=x,y,z} q_{s,\mu \nu} \cdot V^{3,s}_{{\mathrm{H}},\mu \nu}(\vec{r}) + \cdots \right).
\end{eqnarray}
It is noteworthy that the computations of the terms attributed to $V^{1,s}_{\mathrm{H}}(\vec{r})$, $V^{2,s}_{{\mathrm{H}},\mu}(\vec{r})$, and $V^{3,s}_{{\mathrm{H}},\mu \nu}(\vec{r})$ in (\ref{eqn:ono-ewald2D06}) are time-consuming when the periodic boundary condition is used. We will introduce an efficient method to compute these terms in Sect.~\ref{sec:ono-(2D Periodic + 1D Isolated) Boundary Condition}.

\section{Laue representation} 
For 2D periodic materials as shown in Fig.~\ref{tsuka-fig:Figure3}, it is well-known that the Laue representation\cite{PRB59-015609,PRB56-012874} is effective and suitable for describing the systems and their physical quantities. 
In this section, we state the transformations of the 3D real-space equations, i.e., the Kohn-Sham equation for wave functions (\ref{tsuka-eq:KohnShamEquation}) and the Poisson equation for Hartree potential (\ref{tsuka-eq:VHPoissonEquation}), into the Laue representation, which describes physical quantities in wave number space along the $x$ and $y$ directions and in real space along the $z$ direction.

\subsection{Kohn-Sham equation in Laue representation}
Let us start from the continuous 3D Kohn-Sham equation
\begin{equation}
-\frac{1}{2}\left[\frac{\D^{2}}{\D x^{2}}+\frac{\D^{2}}{\D y^{2}}+\frac{\D^{2}}{\D z^{2}}\right]\psi(x,y,z)+v_{\mathrm{eff}}(x,y,z)\psi(x,y,z)=\varepsilon\psi(x,y,z),
\label{tsuka-eq:KohnShamEquationXYZ}
\end{equation}
in which three real-space directions $x$, $y$, and $z$ are explicitly described. 
Because of the periodicity in the $x$ and $y$ directions, the Bloch's theorem can be applied to the wave function $\psi(x,y,z)$ only for the directions of the periodicity, 
\begin{equation}
\psi(x,y,z)=\exp(\I \vec{k}_{\mathrm{||}}\cdot\vec{r}_{\mathrm{||}})u(x,y,z).
\label{tsuka-eq:2DBlochTheorem}
\end{equation}
The Bloch's wave factor $\exp(\I \vec{k}_{\mathrm{||}}\cdot\vec{r}_{\mathrm{||}})$ is determined by the 2D wave number vector $\vec{k}_{\mathrm{||}}=(k_{x},k_{y})$ and 2D position vector $\vec{r}_{\mathrm{||}}=(x,y)$. 
Here, $u(x,y,z)$ is a Bloch function with a periodicity in the $x$ and $y$ directions, which is the same periodicity as the supercell. 
It is easily seen that the periodic function $u(x,y,z)$ can be expanded as a Fourier series only in the $x$ and $y$ directions.
Equation (\ref{tsuka-eq:2DBlochTheorem}) is thus rewritten as 
\begin{equation}
\psi(x,y,z)=\exp(\I \vec{k}_{\mathrm{||}}\cdot\vec{r}_{\mathrm{||}})\sum_{i}\tilde{\psi}_{i}(z)\exp(\I \vec{G}_{\mathrm{||}}^{(i)}\cdot\vec{r}_{\mathrm{||}}),
\label{tsuka-eq:FourierSeriesPsi}
\end{equation}
where $\tilde{\psi}_{i}(z)$ represents the expansion coefficient for $i$th plane wave component and is a continuous function of the $z$ direction. 
Here, $\vec{G}_{\mathrm{||}}^{(i)}$ denotes the $i$th 2D reciprocal vector, and is defined as
\begin{equation}
\vec{G}_{\mathrm{||}}^{(i)}=\left(\frac{2\pi}{L_{x}}i_{x},\frac{2\pi}{L_{y}}i_{y}\right)\quad\mbox{for}\quad i_{x},\,i_{y}=0,\pm1,\pm2,\cdots.
\end{equation}

The effective potential $v_{\mathrm{eff}}(x,y,z)$ in (\ref{tsuka-eq:KohnShamEquationXYZ}) has the periodicity in the $x$ and $y$ directions as well as the system geometry, and thus, is also able to be expanded as a Fourier series in the direction of the periodicity, as
\begin{equation}
v_{\mathrm{eff}}(x,y,z)=\sum_{i}\tilde{v}_{\mathrm{eff}}^{(i)}(z)\exp(\I \vec{G}_{\mathrm{||}}^{(i)}\cdot\vec{r}_{\mathrm{||}}),
\label{tsuka-eq:FourierSeriesVeff}
\end{equation}
where $\tilde{v}_{\mathrm{eff}}^{(i)}(z)$ represents the expansion coefficient for $i$th plane wave component and is also a continuous function of the $z$ direction. 

Substituting (\ref{tsuka-eq:FourierSeriesPsi}) and (\ref{tsuka-eq:FourierSeriesVeff}) into the Kohn-Sham equation (\ref{tsuka-eq:KohnShamEquationXYZ}), one can obtain the Kohn-Sham equation in the Laue representation, 
\begin{multline}
\sum_{i}\left[\frac{1}{2}\left(\vec{k}_{\mathrm{||}}+\vec{G}_{\mathrm{||}}^{(i)}\right)^{2}-\frac{1}{2}\frac{\D^{2}}{\D z^{2}}+\sum_{j}\tilde{v}_{\mathrm{eff}}^{(j)}(z)\exp(\I \vec{G}_{\mathrm{||}}^{(j)}\cdot\vec{r}_{\mathrm{||}})\right]\tilde{\psi}_{i}(z)\exp(\I \vec{G}_{\mathrm{||}}^{(i)}\cdot\vec{r}_{\mathrm{||}})\\
=\varepsilon\sum_{i}\tilde{\psi}_{i}(z)\exp(\I \vec{G}_{\mathrm{||}}^{(i)}\cdot\vec{r}_{\mathrm{||}}).
\end{multline}
By variable transformation, the product of the effective potential and the wave function at the third term in the left-hand side can be further simplified as
\begin{equation}
\sum_{ij}\tilde{v}_{\mathrm{eff}}^{(j)}(z)\exp(\I \vec{G}_{\mathrm{||}}^{(j)}\cdot\vec{r}_{\mathrm{||}})\tilde{\psi}_{i}(z)\exp(\I \vec{G}_{\mathrm{||}}^{(i)}\cdot\vec{r}_{\mathrm{||}})=\sum_{i'j'}\tilde{v}_{\mathrm{eff}}^{(i'-j')}(z)\tilde{\psi}_{j'}(z)\exp(\I \vec{G}_{\mathrm{||}}^{(i')}\cdot\vec{r}_{\mathrm{||}})
\end{equation}
Since the 2D plane wave set $\{\exp(\I \vec{G}_{\mathrm{||}}^{(i)}\cdot\vec{r}_{\mathrm{||}})\}$ forms a complete system, the coefficients to each plane wave component at both sides are equivalent for each $i$. 
Therefore, the wave function expansion coefficient $\tilde{\psi}_{i}(z)$ satisfies the differential equations
\begin{equation}
\frac{1}{2}\left[\left(\vec{k}_{\mathrm{||}}+\vec{G}_{\mathrm{||}}^{(i)}\right)^{2}-\frac{\D^{2}}{\D z^{2}}\right]\tilde{\psi}_{i}(z)+\sum_{j}\tilde{v}_{\mathrm{eff}}^{(i-j)}(z)\tilde{\psi}_{j}(z)=\varepsilon\tilde{\psi}_{i}(z)
\label{tsuka-eq:KohnShamEquationLaue}
\end{equation}
for all $i$.
In practical numerical computation, the expansion coefficients $\tilde{\psi}_{i}(z)$ and $\tilde{v}_{\mathrm{eff}}^{(i-j)}(z)$, which are both the functions of the $z$ direction, have to be discretized. 
For this purpose, we apply the finite-difference approximation (\ref{tsuka-eq:SecondDerivativeFiniteDifferenceApproximation}) to the second derivative of the expansion coefficient function $\tilde{\psi}_{i}(z)$ with respect to the $z$ direction in (\ref{tsuka-eq:KohnShamEquationLaue}). 
Consequently, the Kohn-Sham equation for the wave function coefficient, which is treated in practical numerical computation, is expressed as the linear equations
\begin{equation}
\frac{1}{2}\left(\vec{k}_{\mathrm{||}}+\vec{G}_{\mathrm{||}}^{(i)}\right)^{2}\tilde{\psi}_{i,k}-\frac{1}{2}\sum_{l=-N_{\mathrm{f}}}^{+N_{\mathrm{f}}}\frac{c_{l}}{h_{z}^{2}}\tilde{\psi}_{i,k+l}+\sum_{j}\tilde{v}_{\mathrm{eff}}^{(i-j,k)}\tilde{\psi}_{j,k}=\varepsilon\tilde{\psi}_{i,k}
\label{tsuka-eq:KohnShamEquationLaueRepresentation}
\end{equation}
for all $i$ and $k$. 
As already mentioned above, the 2D materials, on which this article is now focusing, have isolated boundary conditions in the $z$ direction, and hence, the Kohn-Sham Hamiltonian matrix is sparse and band diagonal like as that in the right-hand side of (\ref{tsuka-eq:KohnShamLeftHandSideIsolated}). 

\subsection{Hartree potential in Laue representation}
As already stated in Sect.~\ref{tsuka-sec:DFTKohnShamEquation}, the potential terms have to be determined for constructing Kohn-Sham Hamiltonian and for determining the ground-state electron density. 
Here, we introduce the derivation of the equation for the Hartree potential in the Laue representation, because the Hartree potential is also periodic in the $x$ and $y$ directions and isolated in the $z$ direction. 
The Hartree potential in the real-space representation $v_{\mathrm{H}}(x,y,z)$ is defined as
\begin{equation}
v_{\mathrm{H}}(x,y,z)=\int\frac{\rho(x',y',z')}{|\vec{r}-\vec{r}'|}\D\vec{r}',
\end{equation}
where $\vec{r}=(x,y,z)$ and $\vec{r}'=(x',y',z')$. It is well known that the Hartree potential $v_{\mathrm{H}}(x,y,z)$ satisfies the Poisson equation (\ref{tsuka-eq:VHPoissonEquation}). Because of the geometrical periodicity of the 2D materials, the Hartree potential $v_{\mathrm{H}}(x,y,z)$ and the electron density $\rho(x,y,z)$ are both expressed as Fourier series in the $x$ and $y$ directions, in analogous to the effective potential $v_{\mathrm{eff}}(x,y,z)$ mentioned in the preceding subsection. 
\begin{equation}
v_{\mathrm{H}}(\vec{r})=\sum_{i}\tilde{v}_{\mathrm{H}}^{(i)}(z)\exp(\I\vec{G}_{\mathrm{||}}^{(i)}\cdot\vec{r}_{\mathrm{||}})
\quad
\mbox{and}
\quad
\rho(\vec{r})=\sum_{i}\tilde{\rho}_{i}(z)\exp(\I\vec{G}_{\mathrm{||}}^{(i)}\cdot\vec{r}_{\mathrm{||}})
\label{tsuka-eq:VHRHOFourierTransform}
\end{equation}
Here, $\tilde{v}_{\mathrm{H}}^{(i)}(z)$ and $\tilde{\rho}_{i}(z)$ represent the expansion coefficient functions for the $i$th wave component of the Hartree potential and electron density, respectively. 
Substituting (\ref{tsuka-eq:VHRHOFourierTransform}) into the Poisson equation (\ref{tsuka-eq:VHPoissonEquation}), one can obtain the Poisson equation for the Hartree potential in the Laue representation, as
\begin{equation}
\left[\frac{\D^{2}}{\D z^{2}}-|\vec{G}_{\mathrm{||}}^{(i)}|^{2}\right]\tilde{v}_{\mathrm{H}}^{(i)}(z)=-4\pi\tilde{\rho}_{i}(z)
\end{equation}
for all $i$. 
To discretize the expansion coefficient functions $\tilde{v}_{\mathrm{H}}^{(i)}(z)$ in the $z$ direction, the finite-difference approximation (\ref{tsuka-eq:SecondDerivativeFiniteDifferenceApproximation}) is applied to the second derivative of the expansion coefficient function $\tilde{v}_{\mathrm{H}}^{(i)}(z)$ with respect to the $z$ direction. 
Consequently, the Poisson equation for the Hartree potential expansion coefficients, which is treated in practical numerical computation, is expressed as the linear equations
\begin{equation}
\label{eqn:poissonlinear}
\sum_{l=-N_{\mathrm{f}}}^{+N_{\mathrm{f}}}\frac{c_{l}}{h_{z}^{2}}\tilde{v}_{\mathrm{H}}^{(i,j+l)}-|\vec{G}_{\mathrm{||}}^{(i)}|^{2}\tilde{v}_{\mathrm{H}}^{(i,j)}=-4\pi\tilde{\rho}_{i,j}
\end{equation}
for all $i$ and $j$. 
Now one can see that the matrix operating on the Hartree potential coefficient vector $\tilde{v}_{\mathrm{H}}^{(i,j)}$ is band diagonal as well as the Kohn-Sham Hamiltonian matrix in the Laue representation as seen in (\ref{tsuka-eq:KohnShamEquationLaueRepresentation}). 

To solve (\ref{eqn:poissonlinear}), the boundary value of Hartree potential is required. The details for calculating the boundary value of Hartree potential for 2D materials in practical computations are described in Sect.~\ref{sec:fuzzy cell decomposition and multipole expansion technique}. 

\section{Ewald summation for 2D periodic boundary condition}
\label{sec:ono-(2D Periodic + 1D Isolated) Boundary Condition}
In the case of periodic systems, computations of Coulomb potentials and energies in infinite systems require much computational cost and involve numerical difficulties because the potential of $1/|\vec{r}|$ slowly vanishes at the limit of $|\vec{r}| \rightarrow \infty$. Ewald proposed an efficient method for treating the integrations of Coulomb potentials and energies.\cite{ono-ewald} This method was originally introduced for bulks, in which the periodic boundary condition is imposed on all the directions. In this section, we introduce the extension to the systems in which the periodic boundary condition is imposed in the $x$ and $y$ directions and the isolated boundary condition in the $z$ direction.\cite{ono-prb2005a}

The ionic pseudopotential under the periodic boundary conditions is expressed as
\begin{equation}
\label{eqn:ono-ewald3D01}
v_{\mathrm{ion}}^s(\vec{r}) = \sum_{\vec{P}} \bar{v}^s_{\mathrm{ion}}(\vec{P}+\vec{r}-\vec{R}^s),
\end{equation}
where the sum is performed over real-space lattice vectors $\vec{P}$ of $(n_xL_x,n_yL_y,n_zL_z)$, and $L_x$, $L_y$, and $L_z$ are the lengths of unit cell in the $x$, $y$, and $z$ directions, respectively. Although we assume the use of the pseudopotential proposed by Bachelet et al.\cite{ono-bhs1,ono-bhs2} here, this method is applicable to other types of pseudopotentials, e.g., norm conserving pseudopotential,\cite{PRB43-001993} ultrasoft pseudopotential,\cite{PRB41-007892} and projector augmented-wave (PAW) method.\cite{PRB50-017953} Local components for $s$th atom is given by 
\begin{equation}
\label{eqn:ono-ewald3D00}
\bar{v}^s_{\mathrm{ion}}(\vec{r}) = - \frac{Z_s}{|\vec{r}|} \sum_{i=1,2} C_{s,i} \, \erf (\sqrt{\alpha_{s,i}}\:|\vec{r}|),
\end{equation}
where $C_{s,i}$ ($C_{s,1}+C_{s,2}=1$) and $\alpha_{s,i}$ are the parameters of the pseudopotential and $\erf(x)$ is the error function defined by
\begin{equation}
\erf(x)=\frac{2}{\sqrt{\pi}}\int^x_0 \exp(-t^2)dt.
\end{equation}
Substituting (\ref{eqn:ono-ewald3D00}) into (\ref{eqn:ono-ewald3D01}), we have
\begin{eqnarray}
\label{eqn:ono-ewald3D02}
v_{\mathrm{ion}}^s(\vec{r})&=& - Z_s \sum_{\vec{G}} \frac{2 \pi}{\Omega} \exp \left[\mathrm{i} \vec{G} \cdot (\vec{r}-\vec{R}^s) \right] \int_0^\eta \exp \left(-\frac{|\vec{G}|^2}{4 t^2}\right) \frac{1}{t^3} dt \nonumber \\
&&- Z_s \sum_{\vec{P}} \frac{2}{\sqrt{\pi}} \sum_{i=1,2} C_{s,i} \int_\eta^{\sqrt{\alpha_{s,i}}} \exp (-|\vec{P}+\vec{r}-\vec{R}^s|^2 t^2) dt,
\end{eqnarray}
where $\vec{G}$ are 3D reciprocal vectors of $2\pi(\frac{j_x}{L_x}, \frac{j_y}{L_y}, \frac{j_z}{L_z})$ and $\Omega$ is the volume of unit cell $L_x \times L_y \times L_z$. The Coulomb interactions are divided into a long-range contribution computed in reciprocal space and a short-range sum treated in real space. In typical case for bulks, $\eta$ is chosen to be 0.2 -- 0.7 and the amount of the sums in (\ref{eqn:ono-ewald3D02}) is only $7^3$ -- $11^3$ operations. However, in practical calculations for 2D materials, $\eta$ is chosen so that the potentials and energies rapidly converge with the small numbers of the sums in (\ref{eqn:ono-ewald3D02}) because vacuum is included in the $z$ direction.

\subsection{Ionic pseudopotential}
By using the identity,
\begin{eqnarray}
\lefteqn{\int^1_0 \frac{1}{t^2}\exp\left(-a^2t^2-\frac{b^2}{4t^2}\right)dt} \nonumber \\
&\hspace{0.3cm}=&\frac{\sqrt{\pi}}{2b}\left[\exp(-ab)\,\erfc\left(\frac{b-2a}{2}\right)+\exp(ab)\,\erfc\left(\frac{b+2a}{2}\right)\right] \:\:(b>0),
\end{eqnarray}
the ionic pseudopotential is given by
\begin{eqnarray}
\label{eqn:ono-ewald2D01}
v^s_{\mathrm{ion}}(\vec{r})&=&-\frac{\pi}{S} Z_s \sum_{\left|\vec{G}\right| \ne 0} \frac{1}{|\vec{G}|} \cos \left[ \vec{G} \cdot (\vec{r} - \vec{R}^s) \right] f^+(\vec{G},\vec{r}) \nonumber \\
&&+\frac{2\sqrt{\pi}}{S} Z_s \left[ \frac{1}{\eta} \exp(-|z-R_z^s|^2\eta^2)+\sqrt{\pi} \: |z-R_z^s| \, \erf (|z-R_z^s| \: \eta) \right] \nonumber \\
&&+Z_s \sum_{\vec{P}} \frac{1}{|\vec{\zeta}^s|} \Biggl[ \erf (\eta \: |\vec{\zeta}^s|)-\sum_{i=1,2} C_{s,i} \, \erf (\sqrt{\alpha_{s,i}} \: |\vec{\zeta}^s|) \Biggr],
\end{eqnarray}
where $\vec{G}=2\pi(\frac{j_x}{L_x},\frac{j_y}{L_y},0)$, $\vec{P}=(n_xL_x,n_yL_y,0)$, $S=L_x \times L_y$, $\vec{\zeta}^s=\vec{P}+\vec{r}-\vec{R}^s$, and
\begin{eqnarray}
\label{eqn:ono-ewald2D02}
f^\pm(\vec{G},\vec{r})&=& \exp \bigl[ -|\vec{G}| \: (z-R_z^s) \bigr] \, \erfc \left( \frac{|\vec{G}|-2\eta^2(z-R_z^s)}{2\eta} \right) \nonumber \\
&& \pm \exp \bigl[ |\vec{G}| \: (z-R_z^s) \bigr] \, \erfc \left( \frac{|\vec{G}|+2\eta^2(z-R_z^s)}{2\eta} \right) .
\end{eqnarray}
Here, $\erfc(x)$ is the complementary error function
\begin{equation}
\erfc(x)=1-\erf(x).
\end{equation}

\subsection{Hartree potential}
\label{sec:ono-Hartreepotential}
To determine the boundary values of Hartree potential for the Poisson equation by (\ref{eqn:ono-ewald2D06}), one needs to compute $V^{1,s}_{\mathrm{H}}(\vec{r})$, $V^{2,s}_{{\mathrm{H}},\mu}(\vec{r})$, and $V^{3,s}_{{\mathrm{H}},\mu \nu}(\vec{r})$. It is easily recognized that
\begin{equation}
\label{eqn:ono-ewald2D07}
V^{1,s}_{\mathrm{H}}(\vec{r})=-\lim_{\alpha_{s,i} \rightarrow \infty} \frac{v^s_{\mathrm{ion}}(\vec{r})}{Z_s},
\end{equation}
and therefore,
\begin{eqnarray}
\label{eqn:ono-ewald2D08}
V^{1,s}_{\mathrm{H}}(\vec{r})&=&\frac{\pi}{S} \sum_{\left|\vec{G}\right| \ne 0} \frac{1}{|\vec{G}|} \cos \left[ \vec{G} \cdot (\vec{r}-\vec{R}^s) \right] f^+(\vec{G},\vec{r}) \hspace{3.80cm} \nonumber \\
&&-\frac{2\sqrt{\pi}}{S} \left[ \frac{1}{\eta} \exp(-|z-R_z^s|^2 \: \eta^2)+\sqrt{\pi}|z-R_z^s| \, \erf (|z-R_z^s| \: \eta) \right] \nonumber \\
&&+ \sum_{\vec{P}} \frac{1}{|\vec{\zeta}^s|} \, \erfc (\eta \: |\vec{\zeta}^s|) .
\end{eqnarray}
$V^{2,s}_{{\mathrm{H}},\mu}(\vec{r})$ is obtained by the following relations.
\begin{equation}
\label{eqn:ono-ewald2D03}
V^{1,s}_{\mathrm{H}}(\vec{r}) \equiv \sum_{\vec{P}} \frac{1}{|\vec{\zeta}^s|} ,
\end{equation}
and 
\begin{equation}
\label{eqn:ono-ewald2D04}
V^{2,s}_{{\mathrm{H}},\mu}(\vec{r}) \equiv \sum_{\vec{P}} \frac{\zeta^s_\mu}{|\vec{\zeta}^s|^3} = \frac{\partial}{\partial R^s_{\mu}} V^{1,s}_H(\vec{r}).
\end{equation}
In the case of $\mu=x,y$,
\begin{eqnarray}
\label{eqn:ono-ewald2D09}
V^{2,s}_{{\mathrm{H}},\mu}(\vec{r}) &=& \frac{\pi}{S} \sum_{\left|\vec{G}\right| \ne 0} \frac{G_{\mu}}{|\vec{G}|} \sin \left[ \vec{G} \cdot (\vec{r}-\vec{R}^s) \right] f^+(\vec{G},\vec{r}) \hspace{3.50cm} \nonumber \\
&&+ \sum_{\vec{P}} \left[ \frac{2\eta\exp(-\eta^2 \: |\vec{\zeta}^s|^2)}{\sqrt{\pi} \: |\vec{\zeta}^s|} + \frac{\erfc (\eta \: |\vec{\zeta}^s|)}{|\vec{\zeta}^s|^2} \right] \frac{\zeta^s_\mu}{|\vec{\zeta}^s|} ,
\end{eqnarray}
and in the case of $\mu=z$,
\begin{eqnarray}
\label{eqn:ono-ewald2D10}
V^{2,s}_{{\mathrm{H}},\mu}(\vec{r}) &=& \frac{\pi}{S} \sum_{\left|\vec{G}\right| \ne 0} \cos \left[ \vec{G} \cdot (\vec{r}-\vec{R}^s) \right] f^-(\vec{G},\vec{r}) + \frac{2\pi}{S} \erf (|z-R_z^s| \: \eta) \frac{(z-R_z^s)}{|z-R_z^s|} \nonumber \\
&& + \sum_{\vec{P}} \left[ \frac{2\eta\exp(-\eta^2 \: |\vec{\zeta}^s|^2)}{\sqrt{\pi} \: |\vec{\zeta}^s|} + \frac{\erfc (\eta \: |\vec{\zeta}^s|)}{|\vec{\zeta}^s|^2} \right] \frac{\zeta^s_\mu}{|\vec{\zeta}^s|} .
\end{eqnarray}
There is a similar relation for $V^{3,s}_{{\mathrm{H}},\mu \nu}(\vec{r})$.
\begin{equation}
\label{eqn:ono-ewald2D05}
V^{3,s}_{{\mathrm{H}},\mu \nu}(\vec{r}) \equiv \sum_{\vec{P}} \frac{3\:\zeta^s_\mu \zeta^s_\nu-\delta_{\mu \nu}|\vec{\zeta}^s|^2}{|\vec{\zeta}^s|^5} = \frac{\partial}{\partial R^s_{\mu}}\frac{\partial}{\partial R^s_{\nu}} V^{1,s}_H(\vec{r})
\end{equation}
Thus, in the case of $\mu=x,y$ and $\nu=x,y$,
\begin{eqnarray}
\label{eqn:ono-ewald2D11}
V^{3,s}_{{\mathrm{H}},\mu \nu}(\vec{r}) &=&-\frac{\pi}{S} \sum_{\left|\vec{G}\right| \ne 0} \frac{G_{\mu}G_{\nu}}{|\vec{G}|} \cos \left[ \vec{G} \cdot (\vec{r}-\vec{R}^s) \right] f^+(\vec{G},\vec{r}) \hspace{2.90cm} \nonumber \\
&& + \sum_{\vec{P}} \Biggl[ \left\{ \frac{4 \eta \exp(-\eta^2 |\vec{\zeta}^s|^2)}{\sqrt{\pi}}\left( \eta^2 + |\vec{\zeta}^s|^{-2} \right) + \frac{2\,\erfc (\eta |\vec{\zeta}^s|)}{|\vec{\zeta}^s|^3} \right\} \frac{\zeta^s_\mu \zeta^s_\nu}{|\vec{\zeta}^s|^2} \nonumber \\
&& + \left\{ \frac{2\eta\exp(-\eta^2 \: |\vec{\zeta}^s|^2)}{\sqrt{\pi} \: |\vec{\zeta}^s|} + \frac{\erfc (\eta \: |\vec{\zeta}^s|)}{|\vec{\zeta}^s|^2} \right\} \frac{\zeta^s_\mu \zeta^s_\nu - \delta_{\mu \nu}|\vec{\zeta}^s|^2}{|\vec{\zeta}^s|^3} \Biggr] ,
\end{eqnarray}
in the case of $\mu=x,y$ and $\nu=z$,
\begin{eqnarray}
\label{eqn:ono-ewald2D12}
V^{3,s}_{{\mathrm{H}},\mu \nu}(\vec{r}) &=&\frac{\pi}{S} \sum_{\left|\vec{G}\right| \ne 0} G_{\mu} \sin \left[ \vec{G} \cdot (\vec{r}-\vec{R}^s) \right] f^-(\vec{G},\vec{r}) \hspace{3.45cm} \nonumber \\
&& + \sum_{\vec{P}} \Biggl[ \left\{ \frac{4 \eta \exp(-\eta^2 |\vec{\zeta}^s|^2)}{\sqrt{\pi}}\left( \eta^2 + |\vec{\zeta}^s|^{-2} \right) + \frac{2\,\erfc (\eta |\vec{\zeta}^s|)}{|\vec{\zeta}^s|^3} \right\} \frac{\zeta^s_\mu \zeta^s_\nu}{|\vec{\zeta}^s|^2} \nonumber \\
&& + \left\{ \frac{2\eta\exp(-\eta^2 \: |\vec{\zeta}^s|^2)}{\sqrt{\pi} \: |\vec{\zeta}^s|} + \frac{\erfc (\eta \: |\vec{\zeta}^s|)}{|\vec{\zeta}^s|^2} \right\} \frac{\zeta^s_\mu \zeta^s_\nu - \delta_{\mu \nu}|\vec{\zeta}^s|^2}{|\vec{\zeta}^s|^3} \Biggr] ,
\end{eqnarray}
and in the case of $\mu=\nu=z$,
\begin{eqnarray}
\label{eqn:ono-ewald2D13}
V^{3,s}_{{\mathrm{H}},\mu \nu}(\vec{r}) &=&\frac{\pi}{S} \sum_{\left|\vec{G}\right| \ne 0} \cos \left[ \vec{G} \cdot (\vec{r}-\vec{R}^s) \right] \hspace{5.60cm} \nonumber \\
&& \times \left[ \left|\vec{G}\right| f^+(\vec{G},\vec{r}) - \frac{4\eta}{\sqrt{\pi}} \exp \left\{ -\frac{|\vec{G}|^2+4(z-R_z^s)^2\eta^4}{4\eta^2} \right\} \right] \nonumber \\
&&-\frac{4\sqrt{\pi}\eta}{S}\exp(-|z-R_z^s|^2\eta^2) \nonumber \\
&& + \sum_{\vec{P}} \Biggl[ \left\{ \frac{4 \eta \exp(-\eta^2 |\vec{\zeta}^s|^2)}{\sqrt{\pi}}\left( \eta^2 + |\vec{\zeta}^s|^{-2} \right) + \frac{2\,\erfc (\eta |\vec{\zeta}^s|)}{|\vec{\zeta}^s|^3} \right\} \frac{\zeta^s_\mu \zeta^s_\nu}{|\vec{\zeta}^s|^2} \nonumber \\
&& + \left\{ \frac{2\eta\exp(-\eta^2 \: |\vec{\zeta}^s|^2)}{\sqrt{\pi} \: |\vec{\zeta}^s|} + \frac{\erfc (\eta \: |\vec{\zeta}^s|)}{|\vec{\zeta}^s|^2} \right\} \frac{\zeta^s_\mu \zeta^s_\nu - \delta_{\mu \nu}|\vec{\zeta}^s|^2}{|\vec{\zeta}^s|^3} \Biggr] .
\end{eqnarray}

\subsection{Coulomb energy among the nuclei}
Coulomb energy among the nuclei is
\begin{eqnarray}
\label{eqn:ono-ewald2D14}
\gamma_E &=& \frac{\pi}{2S} \sum_{s,s'} Z_s Z_{s'} \sum_{\left|\vec{G}\right| \ne 0} \frac{1}{|\vec{G}|} \cos \left[ \vec{G} \cdot (\vec{R}^{s'} - \vec{R}^s) \right] f^+(\vec{G},\vec{R}^{s'}) \nonumber \\
&&-\frac{\sqrt{\pi}}{S} \sum_{s,s'} Z_s Z_{s'} \left[ \frac{1}{\eta} \exp (-|R_z^{s'}-R_z^s|^2 \eta^2) + \sqrt{\pi} \: |R_z^{s'}-R_z^s| \, \erf (|R_z^{s'}-R_z^s| \eta) \right] \nonumber \\
&& - \sum_{s,s'} Z_s Z_{s'} \delta_{ss'} \frac{\eta}{\sqrt{\pi}} +\frac{1}{2} \sum_{\vec{P},s,s'} \!\!\!\mbox{}^\prime Z_s Z_{s'} \frac{\erfc (\eta \: |\vec{\xi}^{s,s'}|)}{|\vec{\xi}^{s,s'}|} ,
\end{eqnarray}
where $\vec{\xi}^{s,s'}=\vec{P}+\vec{R}^{s'}-\vec{R}^s$.

A nucleus does not interact with its own Coulomb charge, so that the $\vec{P}=0$ term must be omitted from the real-space summation when $s=s'$. The prime in the last summation in (\ref{eqn:ono-ewald2D14}) means that $|\vec{\xi}^{s,s'}|=0$ is omitted.

\section{Electronic band structure for 2D materials}
\label{sec:ono-3D Periodic Boundary Condition: Crystals}
In order to ensure the efficiency of the real-space finite-difference method, we have calculated the electronic band structure of the several 2D materials, graphene, silicene, and germanene. These systems are usually treated with the plane-wave basis set using slab models. However, we can treat truly 2D model by virtue of the advantage of the real-space finite-difference method. The computational conditions are as follows: the coarse-grid spacings $h_i$ ($i=x$, $y$, and $z$) are set at $\sim$ 0.30 $a_{\mathrm{B}}$, $\sim$ 0.40 $a_{\mathrm{B}}$, and $\sim$ 0.40 $a_{\mathrm{B}}$, for graphene, silicene, and germanene, respectively. The dense-grid spacing of the double-grid method\cite{ PRB72-085115,ono-tsdg1,ono-tsdg3} is fixed at $h_i/2$. The {\it k}-space integrations are performed with 142 {\bf k} points in the irreducible Brillouin zone. The nine-point finite-difference formula ($N_\mathrm{f}=4$) and the local density approximation (LDA)\cite{ono-lda} are employed. The PAW method\cite{PRB50-017953} is used for interaction between electrons and nuclei. Coulomb potentials and energies in infinite systems are computed using the formulae given in Sect.~\ref{sec:ono-(2D Periodic + 1D Isolated) Boundary Condition}. After the structural optimization, we found that silicene and germanene are slightly buckled while graphene is planar. Figure~\ref{fig:ono-band} shows the electronic band structures. Although the bands attributed to $\sigma$ orbitals approach to the Fermi level as the atomic number increases, the particular feature of $sp^3$ bonding network is observed at the Fermi level. Thus, silicene and germanene are also promising 2D materials as well as graphene.

\section{First-principles studies on MoS$_2$ monolayer}
\label{sec:egami1}
Recently, considerable interest has been focused on TMD compounds due to their great potential as a complementary material to graphene. TMD compounds have the chemical formula TX$_2$, where T represents a transition metal from group IV, V, or VI, and X represents a chalcogen (S, Se, or Te). The most remarkable feature of TX$_2$ compounds is their highly anisotropic layered structure. While TX$_2$ layers are linked to each other by weak van-der-Waals-type forces, each TX$_2$ layer is composed of covalently bonded X-T-X sandwiches. In other words, in each layer, a transition-metal plane is enclosed within two chalcogen planes with a trigonal prismatic configuration to form a hexagonal crystal structure. Furthermore, all TX$_2$ compounds are indirect band gap semiconductors.\cite{WYLiang1986,AAruchamy1992,ThBoeker2001,AKlein2001}
Owing to the weak interlayer interactions, an isolated TX$_2$ monolayer can readily be extracted mechanically using an adhesive tape,\cite{RFFrindt1966} chemically by lithium-based intercalation,\cite{PJoensen1986} or by employing other techniques\cite{VNicolosi2013} analogous to the exfoliation of graphene. In this case, the exposed TX$_2$ surfaces are relatively inert without any intrinsic dangling bonds since no covalent bonds are broken during the exfoliation process.

MoS$_2$ (molybdenum disulfide) is a typical example of a several-layered TMD compound (Fig.~\ref{fig:egami01}). For the last several decades, bulk MoS$_2$ has been extensively investigated and attracted attention due to its unique electronic, optical, and catalytic properties\cite{AAruchamy1992,ThBoeker2001,GKline1982,JDFuhr1999,ZHChi2014} as well as its importance for dry lubrication.\cite{LRapoport1997} Furthermore, the intriguing properties of the MoS$_2$ monolayer, which are expected to be distinct from those of bulk MoS$_2$, are considered to have broad application prospects in next-generation nanodevices, and thus, have been arousing tremendous interest recently.
In particular, the MoS$_2$ monolayer has an intrinsic direct band gap\cite{SLebegue2009,ASplendiani2010,KFMak2010,TKorn2011,AKuc2011,ARamasubramaniam2012,DXiao2012,AMSanchez2013,WJin2013,DLiu2013,MCalandra2013} that differs from that of bulk MoS$_2$, which is an indirect gap semiconductor.\cite{ThBoeker2001,KKKam1982} Therefore, the 2D MoS$_2$ monolayer is more attractive than graphene from the viewpoint of the applications to nano-electronic devices such as high-speed and low-power-consumption FETs, where a high on/off current ratio is required. Simultaneously, this material has shown to be promising for applications in optoelectronics, energy storage, and energy harvesting fields.\cite{ASplendiani2010,JXiao2010,BRadisavljevic2011Nature,KFMak2012,WJYu2013}
In the development of such novel electronic devices, the unique characteristics of the MoS$_2$ monolayer have motivated many researchers to eagerly seek techniques to tune its outstanding electronic, chemical, and physical properties in a controllable manner by either introducing lattice defects (e.g., introducing vacancies, or substituting and/or adsorbing impurities),\cite{DLiu2013,JHe2010,KDolui2013,ZHuang2013,XLin2014,HSSen2014,HZheng2014,JChang2014,FMehmood2014,JQi2014,XDLi2015} or imposing external forces (e.g., mechanical strain, or electronic and/or magnetic fields).\cite{HZheng2014,JQi2014,SMTabatabaei2013,JQi2013} A few layered structures\cite{JKEllis2011,NTCuong2014,HPeelaers2014,SBhattacharyya2014} are also attractive, which include hetero-layered structures with a MoS$_2$ monolayer adsorbed on metals,\cite{IPopov2012,WASaidi2014} semiconductor oxide surfaces,\cite{SWHan2011,HJSung2014,WLScopel2015} and other 2D materials.\cite{SWHan2011,EScalise2014,RGillen2014}

In this section, we introduce recent theoretical works to elucidate and control the prominent properties of the MoS$_2$ monolayer using first-principles calculations within the framework of the density functional theory. Here we focus on the modulation of the electronic and magnetic properties of the MoS$_2$ monolayer by the lattice defects mentioned above.

To investigate the electronic properties of the bulk MoS$_2$, many studies on the electronic band structure have been performed using experimental\cite{ThBoeker2001,KFMak2010,WJin2013,TKomesu2014,SKMahatha2014} and theoretical\cite{AMSanchez2013,SBhattacharyya2014,SKMahatha2014,TCheiwchanchamnangij2012,BHu2015} methods. In these studies, the 3D bulk MoS$_2$ crystal is reported to be an indirect semiconductor with an indirect band gap of $\sim{1.3}$~eV and a direct band gap of $\sim{1.8}$~eV. Those arise from the presence of two minima in the conduction band (CB) at different $k$-points in the hexagonal first-Brillouin zone. Here, one minimum with a higher energy is directly above the valence band (VB) maximum at the K point and the other with a lower energy is located between the K and ${\rm \Gamma}$ points. As the thickness of the MoS$_2$ crystal is decreased to a few atomic layers, the indirect band gap becomes wider while the direct band gap is insensitive to the number of layers. Finally, experimental studies find an indirect-to-direct band gap transition in the pristine 2D monolayer form, which has a direct band gap of $\sim{1.9}$~eV.\cite{ASplendiani2010,KFMak2010,TKorn2011,WJin2013}

\subsection{Electronic band structure of pristine MoS$_2$ monolayer\label{ssec:egami1-1}}
To reveal the electronic band structure and band gap of the MoS$_2$ monolayer, several first-principles studies have also been performed.\cite{SLebegue2009,AKuc2011,ARamasubramaniam2012,DXiao2012,AMSanchez2013,DLiu2013,JHe2010,KDolui2013,ZHuang2013,XLin2014,JChang2014,JKEllis2011,SWHan2011,HJSung2014,TCheiwchanchamnangij2012,TLi2007,CAtaca2011,HPKomsa2012,AKumar2012,DYQiu2013,SKC2014}
Figure~\ref{fig:egami04} represents the electronic band structure of the MoS$_2$ monolayer. The electronic band structure is calculated based on the real-space finite-difference formalism using the nine-point finite-difference formula ($N_{\mathrm{f}}=4$), the coarse-grid spacings of 0.33~$a_\mathrm{B}$, a 144 mesh for the \emph{k}-space integration in the first-Brillouin zone, the PAW method,\cite{PRB50-017953} and the Perdew-Burke-Ernzerhof generalized gradient approximation.\cite{JPPerdew1996} The atomic configrations are set to be $a=6.00~a_{\mathrm{B}}$ and $c=5.93~a_{\mathrm{B}}$ in Fig.~\ref{fig:egami01}. As the result, the direct band gap of 1.73~eV is obtained in our calculation.
However, the value of the band gap varies greatly according to the lattice constant of the models and the procedures employed to treat the exchange-correlation potential and the ionic pseudopotential. 
For instance, the computed band gap is sensitive to the exchange-correlation functional adopted in the Kohn-Sham equation. While the computed band gap is $1.60-1.90$~eV\cite{SLebegue2009,AKuc2011,ARamasubramaniam2012,AMSanchez2013,JHe2010,KDolui2013,ZHuang2013,XLin2014,JChang2014,SWHan2011,HJSung2014,TLi2007,AKumar2012,DYQiu2013,SKC2014} using the LDA\cite{ono-lda} and the generalized gradient approximation,\cite{JPPerdew1996,JPPerdew1992} the gap is $2.05-2.30$~eV\cite{ARamasubramaniam2012,KDolui2013,JKEllis2011,CAtaca2011} when estimated by employing the functional of Heyd, Scuseria, and Ernzerhof,\cite{JHeyd2003,JHeyd2006} and $2.41 - 2.97$~eV\cite{SLebegue2009,ARamasubramaniam2012,AMSanchez2013,TCheiwchanchamnangij2012,CAtaca2011,HPKomsa2012,DYQiu2013} when estimated by the GW approximation,\cite{LHedin1965} respectively.
Moreover, the band splitting of $\sim{150}$~meV has been reported to be induced by the strong spin-orbit coupling around the K point due to broken inversion symmetry.\cite{ARamasubramaniam2012,DXiao2012,AMSanchez2013,HJSung2014,TCheiwchanchamnangij2012,ZYZhu2011,AKormanyos2013,JHe2014} This characteristic makes MoS$_2$ interesting for spin-physics exploration.

\subsection{Vacancy defect\label{ssec:egami1-2}}
During the experimental fabrication of single-layered MoS$_2$ using epitaxial growth, chemical vapor deposition,\cite{WLi2014} or mechanical exfoliation techniques,\cite{RFFrindt1966,KSNovoselov2005} vacancy defects are observed due to the imperfection of the growth or exfoliation process. Thus, several dangling bonds appear in the MoS$_2$ monolayer and largely affect the electronic band structure around the band gap, where the VB (CB) edge is mainly constituted from the hybridization of Mo $4d$ and S $3p$ orbitals (Mo $4d$ orbitals).
In order to examine possible defect structures, first-principles calculations were performed\cite{SKC2014} using computational models including mono-S vacancy (V$_{\rm S}$), di-S vacancy (V$_{\rm S_2}$), tri-S vacancy (V$_{\rm S_3}$), and mono-Mo vacancy (V$_{\rm Mo}$) as shown in Fig.~\ref{fig:egami02}.
When V$_{\rm S}$ is introduced into the MoS$_2$ monolayer, a defect state close to the CB edge exists, which arises from the dangling bond of Mo $4d$ orbitals. In addition, the reduction of the hybridization between Mo $4d$ and S $3sp$ orbitals induce a shallow state change close to the VB maximum. These defect states are localized around Mo atoms adjacent to V$_{\rm S}$.
While it is energetically difficult to form V$_{\rm S_2}$, V$_{\rm S_3}$, or V$_{\rm Mo}$, once they are formed, the defect states are populated in the upper half of the band gap and are extended up to the mid-gap region. This means that it is very important to establish schemes to suppress the forming of vacancy defects and to terminate the dangling bonds.

On the other hand, vacancy defects can be utilized to induce the magnetic properties in the MoS$_2$ monolayer.
Zheng et al.\cite{HZheng2014} investigated the MoS$_2$ monolayer including four types of vacancies in a supercell consisting of $6 \times 6$ 2D unit cells: V$_{\rm S}$, V$_{\rm S_2}$, the vacancy complex of Mo atom and three nearby S atoms(V$_{\rm MoS_3}$), and the vacancy complex of Mo atom and nearby three top-and-bottom S atom pairs (V$_{\rm MoS_6}$) (Fig.~\ref{fig:egami02}).
In the case of \emph{non-relaxed} configurations, the dangling bonds are found to induce magnetic moment in the MoS$_{2}$ monolayers. On the other hand, in the \emph{fully relaxed} configurations, V$_{\rm S}$-, V$_{\rm S_2}$-, and V$_{\rm MoS_3}$-doped systems do not exhibit any magnetic properties because the unsaturated spin electrons in the S and Mo atoms around these vacancy defects are paired, owing to variations in the bonding circumstances of S--Mo covalent bonds and Mo--Mo metallic bonds. On the contrary, there remain several localized non-bonding $4d$ orbitals of the six Mo atoms around the V$_{\rm MoS_6}$ defect, which induce a large magnetic moment of 6.0~$\mu_{\mathrm{B}}$ per $6 \times 6$ unit cells.

In addition, an effective scheme to control the magnetic properties involves the application of a mechanical strain to induce geometrical distortion. It was reported that the tensile strain can induce the transitions in the magnetic properties of the MoS$_2$ monolayer from non-magnetic to ferromagnetic (FM).\cite{HZheng2014} Such strain engineering is also a useful approach for modulating the electronic and mechanical properties of the MoS$_2$ monolayer.
The magnetic properties of the vacancy-doped MoS$_2$ monolayer under the application of strain have also been examined. Equibiaxial tensile strains larger than $\epsilon_{\rm biax} = 9~\%$ can lead to the formation of a magnetic moment of 2.0 (5.5)~$\mu_{\mathrm{B}}$ per $6 \times 6$ unit cells for a V$_{\rm S}$- (V$_{\rm S_2}$-) doped system, where $\epsilon_{\rm biax} = (\ell-\ell_0)/\ell_0$ with the constrained and unstrained lattice constants being $\ell$ and $\ell_0$, respectively. This is caused by the breaking of the Mo--Mo bonds under the relatively large strain, which results in the formation of localized non-bonding $4d$ electrons of Mo atoms around the vacancies. Similarly, for a V$_{\rm MoS_3}$-doped system, strains larger than $\epsilon_{\rm biax} = 10~\%$ give rise to the formation of such $4d$ orbitals.
On the other hand, under the compressive strain corresponding to $\epsilon_{\rm biax} < 0$, the magnetic moment decreases because the Mo--Mo metallic bond is strong and leads to a decrease in the non-bonding $4d$ electrons. Thus, the magnetic moment for a V$_{\rm MoS_6}$-doped system can be tuned between 0.0 and 12.0~$\mu_{\mathrm{B}}$ by varying the strain $\epsilon_{\rm biax}$ from $-12~\%$ to $9~\%$. Consequently, V$_{\rm S}$-, V$_{\rm S_2}$-, and V$_{\rm MoS_3}$-doped systems exhibit magnetic properties by imposing the tensile strain and the magnetic moment for V$_{\rm MoS_6}$-doped systems can be tuned by the strain.

\subsection{Adsorption of atom and molecule\label{ssec:egami1-3}}
In order to modulate the electronic band structures, the adsorption of atoms on the MoS$_2$ monolayer surface is an effectual approach, which is analogous to the strategy for graphene.
Metallic, semi-metallic, or semiconducting behavior will occur depending on the adatom type.\cite{JHe2010,KDolui2013,ZHuang2013,JChang2014} Moreover, the magnetic properties can be introduced by adatoms as well as vacancy defects,\cite{HZheng2014} and thus, a scheme to control the magnetism is highly desirable.
For example, Huang et al.\cite{ZHuang2013} investigated the modulation of the electronic and magnetic properties of a Fe atom adsorbed on the MoS$_2$ monolayer composed of $4 \times 4$ unit cells.
On the MoS$_2$ surface, there are four-type adsorption sites as represented in Fig.~\ref{fig:egami03}: an on-top site above a S atom, an on-top site above a Mo atom, a bridge site between S and Mo atoms, and a hollow site located at the center of a hexagonal ring consisting of three S and Mo atoms.
According to the total energy calculations, for the Fe adatom, the on-top site above the Mo atom is the most favorable. In this case, the Fe adatom is covalently bonded to three neighboring S atoms with a charge transfer of 0.82 electrons from the Fe adatom to the S atoms.
As the result, the local magnetic moment of the Fe adatom is reduced to 1.9~$\mu_{\mathrm{B}}$ from 4.0 $\mu_{\mathrm{B}}$ being the magnetic moment of a free-standing Fe atom. When the free Fe atom adsorbs on the MoS$_2$ surface, electrons transfer from the Fe $4s$ orbital to not only the MoS$_2$ but also the Fe $3d$ and $4p$ orbitals. Therefore, the number of unpaired electrons of the $3d$ orbitals decreases and the local magnetic moment is reduced to 1.9~$\mu_{\mathrm{B}}$.
In addition, the adsorption of the Fe atom induces impurity states within the band gap of the pristine MoS$_2$ monolayer, which are formed from minority-spin electronic states, and thus the band gap becomes smaller than that of the pristine system.

The effects of other adatoms on the electronic and magnetic properties of the MoS$_2$ monolayer have been extensively investigated. He et al.,\cite{JHe2010} Dolui et al.,\cite{KDolui2013} Sen et al.,\cite{HSSen2014} Chang et al.,\cite{JChang2014} and Li et al.\cite{XDLi2015} have reported systematic studies on MoS$_2$ monolayers with several adatom.
In the adsorption energy calculations for non-metal adatoms, energetically favorable absorption sites differ according to the species of adatom. For example, H, N, O, and F adatoms favor adsorption on the on-top site above the S atom, while the on-top site of the Mo atom is the most stable for B and C adatoms.\cite{JHe2010} The hollow site is one of the favorable adsorption sites for the graphene system. However, the size of the hexagonal ring is so large that the hollow site on the MoS$_2$ surface is not energetically stable, with the exception of some transition-metal adatoms (Sc, Ti, Mn, and Ag).\cite{JChang2014}
Similar to the MoS$_{2}$ monolayer with vacancy defects, the spin-polarized state is observed. More specifically, magnetic moments of 1.0, 1.0, 2.0, 1.0, and 1.0~$\mu_{\mathrm{B}}$ are obtained per single adatom in $4 \times 4$ unit cells for the H-, B-, C-, N-, and F-absorbed MoS$_2$ systems, respectively.\cite{JHe2010} Although the magnetic properties can be induced by these impurities, the contributions of the adatoms to the total magnetic moments are not very large. In addition, the spin polarizations of the S $3p$ and Mo $4d$ electrons around the adatoms are induced by the localized and spin-polarized $2p$ orbitals ($1s$ of H atom) of the adatoms.\cite{JHe2010}
In contrast, the O-absorbed system is non-magnetic with zero total magnetic moment. Here, the $2p$ orbitals of O overlap with the $3p$ orbitals of S atom to form a strong bond, and the charge transfer occurs only between the O adatom and the neighboring S atom, where the S atom donates electrons to O and becomes positively charged and none of the other atoms is significantly affected.\cite{HSSen2014}

It is also seen from the density of states that the spin-polarized impurity state in the H- (N-) absorbed system is located near the CB minimum (VB maximum) and the system is treated as an $n$-type ($p$-type) semiconductor. On the other hand, in the case of the B- , C-, and F-adsorbed systems, the impurity states are found in the mid-gap and the magnetic moments of the system originate according to the number of electrons occupied in these states.
Moreover, the impurity states caused by hybridization of the $2p$ states of adatoms ($1s$ of H atom) and the valence states of the host species at the neighboring position give rise to the reduction of the band gap. In particular, for the H- (F-) adsorbed system, a large spatial extension of the spin density is observed since the H $1s$ (F $2p$) state additionally overlaps with the second-nearest S $3s$ states and the third-nearest Mo $4d$ states, and weak anti-FM coupling is observed.\cite{JHe2010}

When alkali-metal atoms absorb on the on-top site above the Mo atom with ionic bonding, the MoS$_2$ monolayer possesses the characteristics of an $n$-type semiconductor since the Fermi level is shifted into the CB while there are no crucial modulations in the band gap and the electronic band structures near the band edge.\cite{JChang2014,XDLi2015} On the other hand, halogen adatoms and transition-metal adatoms are effective dopants for the MoS$_2$ monolayer with the exception of Sc and Pd adatoms, where the states created near the band edges might serve as donor- and acceptor-like states, respectively, in a highly dielectric environment.\cite{JChang2014}

In contrast to the adatom-chemisorbed systems, the studies on the physisorption of the chemical compounds on the MoS$_2$ surface have also been performed by Mehmood and Pachter,\cite{FMehmood2014} where non-empirical van-der-Waals corrections are included. In the studies, the charge transfer upon the adsorption of chemical compounds is examined using the Bader charge analysis.\cite{RFWBader1990} The adsorption energy of molecules on the MoS$_2$ surface is comparable to that on the graphene surface. However, accumulation and depletion of electrons for almost all adsorbates are observed due to the charge transfer between the molecules and the free-standing MoS$_2$ monolayer or that with the SiO$_2$ substrate, which is slightly larger than that between molecules and the graphene surface. This result indicates the intrinsic potential of the MoS$_2$ monolayer for a chemical-sensing application.

\subsection{Substitutional doping\label{ssec:egami1-4}}
Dolui et al.\cite{KDolui2013} also examined the substitutional doping at the Mo and S sites.
Such substitutional doping is an alternative scheme to modulate the electronic and magnetic properties.
Qi et al.,\cite{JQi2014} Lin et al.,\cite{XLin2014} and Ramasubramaniam et al.\cite{ARamasubramaniam2012} examined MoS$_2$ monolayers with Mn, Fe, Co, and Zn atoms substituting for a Mo atom, and observed that they exhibit magnetic properties.
For the system with $n$-type (halogen atoms) and $p$-type (group V elements) doping at a S site, most dopants form localized and spin-polarized states within the band gap of the MoS$_2$ monolayer. The dopants exhibit the ground state with a magnetic moment of 1.0~$\mu_{\mathrm{B}}$, while I- and As-doped systems display non-magnetic behavior.
On the other hand, when a Mo atom is substituted by other transition-metal elements with a different number of $4d$ electrons, the dopants with more $d$ electrons than the Mo atom tend to create the donor states deep inside the band gap and give rise to large magnetic moments. In contrast, Nb, Zr, and Y with less $4d$ electrons as dopants form the defect state near the VB maximum and do not induce the magnetic moment.\cite{KDolui2013} 
Such systematic investigations are remarkably useful to understand the fundamental properties and to explore the possibilities for applications in several fields.

As mentioned above, mechanical strain is utilized to tune the magnetic properties of the MoS$_2$ monolayer with adatoms and vacancy defects. This strategy can also be an effective approach to control the magnetic properties of the substitutional doping system. 
Qi et al.\cite{JQi2014} researched the controllability of the magnetic properties in a Mn-doped MoS$_2$ monolayer by a biaxial strain. In the unstrained Mn-doped MoS$_2$ monolayer, where one Mo atom is replaced by a Mn atom, the overall magnetic moment per $4 \times 4$ unit cells is 1.0~$\mu_{\mathrm{B}}$ corresponding to the single excess $d$ electron provided by the Mn atom.
As the tensile strain is larger, the interatomic distance between Mn and S atoms increases gradually, and thus, the covalent bonding between the atoms becomes weaker and a metastable state with a magnetic moment of 3.0~$\mu_{\mathrm{B}}$ is observed. In this case, the local atomic magnetic moments slowly increase around the doped Mn atoms up to 4.3~$\mu_{\mathrm{B}}$, while the surrounding Mo and S atoms display total magnetic moments of $-0.5~\mu_{\mathrm{B}}$ and $-0.8~\mu_{\mathrm{B}}$, respectively. Then, the energy difference between the state with 3.0~$\mu_{\mathrm{B}}$ and 1.0~$\mu_{\mathrm{B}}$ decreases gradually. Finally, the state with 3.0~$\mu_{\mathrm{B}}$ becomes the magnetic ground state at strains larger than $4.5~\%$.
Furthermore, the energy difference between the FM state and anti-FM state is examined, and it is found that the FM state is much more stable and the stability is insensitive to the strain.
Lin et al.\cite{XLin2014} reported that the electronic and magnetic properties of the MoS$_2$ monolayer substitutionally doped with Mn, Fe, and Co can be tuned depending on their possible charge states.

\section{Summary and outlook\label{sec:conclusion}}
In this review, we have presented the theoretical procedures based on the real-space finite-difference formalism to compute the electronic and magnetic properties of 2D materials with a high accuracy and efficiency.
The practical formulae to obtain the Kohn--Sham effective potential and solve the Kohn--Sham equation were derived under the 2D periodic boundary condition. 
To demonstrate the performance of the proposed procedures, we calculated the energy band structures of graphene, silicene, and germanene. Silicene and germanene have the particular feature of $sp^3$ bonding network at the Fermi level, and thus, they are promising 2D materials as well as graphene.
Moreover, the real-space finite-difference formalism is suitable for massively parallel computers consisting of thousands of cores. Therefore, the proposed procedures are remarkably powerful tools for designing next-generation electronic devices by employing large scale 2D models.

The band-gap engineering and induction of magnetic properties using lattice vacancies, adatoms, substitutional defects and mechanical strain are important techniques for the design and development of new semiconductor materials and high-performance devices based on 2D materials. As introduced in this article, most of the intensive investigations on 2D MoS$_2$ thin films have been performed in this half decade, and more investigations are still in progress and are expected to provide fruitful results for MoS$_2$ and other 2D materials. If the techniques to easily control and tune the electronic and magnetic properties of 2D materials according to the requirements for each application field can be successfully established, they will lead to innovation and a great impact for both fundamental physics and engineering applications to the electronics, spintronics and optical device fields. Our theoretical procedures can facilitate the studies on 2D materials.

\section*{Acknowledgments}
The authors are grateful to Professor Dr.~S.~Bl{\"u}gel of Forschungszentrum J{\"u}lich for valuable counsels. This research was partially supported by the Computational Materials Science Initiative (CMSI) from the Ministry of Education, Culture, Sports, Science and Technology, Japan. The numerical calculation was carried out using the computer facilities of the Institute for Solid State Physics at the University of Tokyo and Center for Computational Sciences at University of Tsukuba.
%

\newpage
\begin{table}[p]
\caption{Real-space finite-difference coefficients of the second derivative $c_{l}$ for approximation orders of $N_{\mathrm{f}}=1,\cdots,8$.}
\label{tsuka-tbl:Table1}
{
\newlength\cellheight
\settoheight{\cellheight}{$\displaystyle\frac{1}{2}$}
\newcolumntype{C}{>{\hfil$\displaystyle}c<{$\hfil}}
\begin{tabular}{CCCCCCCCCC}
\hline
 N_{\mathrm{f}} & c_{0} & c_{\pm1} & c_{\pm2} & c_{\pm3} & c_{\pm4} & c_{\pm5} & c_{\pm6} & c_{\pm7} & c_{\pm8} \\
\hline
 1 & -2 & 1\rule{0pt}{\cellheight} \\[2ex]
 2 & -\frac{5}{2} & \frac{4}{3} & -\frac{1}{12} \\[2ex]
 3 & -\frac{49}{18} & \frac{3}{2} & -\frac{3}{20} & \frac{1}{90} \\[2ex]
 4 & -\frac{205}{72} & \frac{8}{5} & -\frac{1}{5} & \frac{8}{315} & -\frac{1}{560} \\[2ex]
 5 & -\frac{5269}{1800} & \frac{5}{3} & -\frac{5}{21} & \frac{5}{126} & -\frac{5}{1008} & \frac{1}{3150}  \\[2ex]
 6 & -\frac{5369}{1800} & \frac{12}{7} & -\frac{15}{56} & \frac{10}{189} & -\frac{1}{112} & \frac{2}{1925} & -\frac{1}{16632} \\[2ex]
 7 & -\frac{266681}{88200} & \frac{7}{4} & -\frac{7}{24} & \frac{7}{108} & -\frac{7}{528} & \frac{7}{3300} & -\frac{7}{30888} & \frac{1}{84084} \\[2ex]
 8 & -\frac{1077749}{352800} & \frac{16}{9} & -\frac{14}{45} & \frac{112}{1485} & -\frac{7}{396} & \frac{112}{32175} & -\frac{2}{3861} & \frac{16}{315315} & -\frac{1}{411840} \\[2ex]
\hline
\end{tabular}
}
\end{table}

\newpage
\begin{figure}[p]
\includegraphics{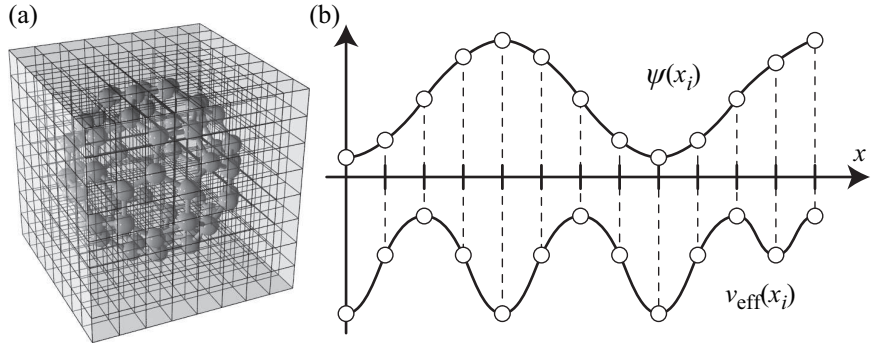}
\caption{Schematic representations of the real-space finite-difference formalism. {\bf(a)} illustrates the discretization of the 3D real-space domain containing a C$_{60}$ molecule. The cross points of the black lines represent the grid points. {\bf(b)} draws the sampling of wave function $\psi(x)$ and effective potential $v_{\mathrm{eff}}(x)$ on the discretized grid points $x_{i}$ along the $x$ direction.}
\label{tsuka-fig:Figure1}
\end{figure}

\newpage
\begin{figure}[p]
\includegraphics{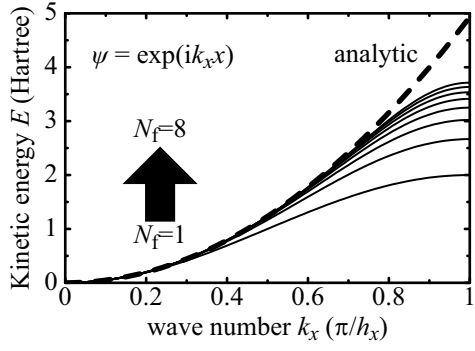}
\caption{Error estimation of the finite-difference approximation for the kinetic energy. The solid curves show the energy dispersion relations of a plane wave $\exp(\I k_{x}x)$ evaluated by using the finite-difference formula $(\ref{tsuka-eq:SecondDerivativeFiniteDifferenceApproximation})$ with the approximation orders of $N_{\mathrm{f}}=1,\cdots,8$ and the grid spacing $h_{x}=1a_{\mathrm{B}}$. The dashed curve shows the analytical energy dispersion relation of $\frac{1}{2}k_{x}^{2}$.}
\label{tsuka-fig:Figure2}
\end{figure}

\newpage
\begin{figure}[p]
\includegraphics[scale=0.75]{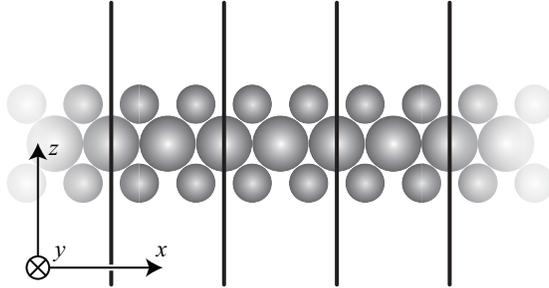}
\caption{Schematic representation of MoS$_{2}$ monolayer, one of the functional 2D materials. Large and small spheres represent Mo and S atoms, respectively. In the $x$ and $y$ directions parallel to the film periodicity exists, while in the $z$ direction perpendicular to the film the system is isolated.}
\label{tsuka-fig:Figure3}
\end{figure}

\newpage
\begin{figure}[p]
\begin{center}
\includegraphics{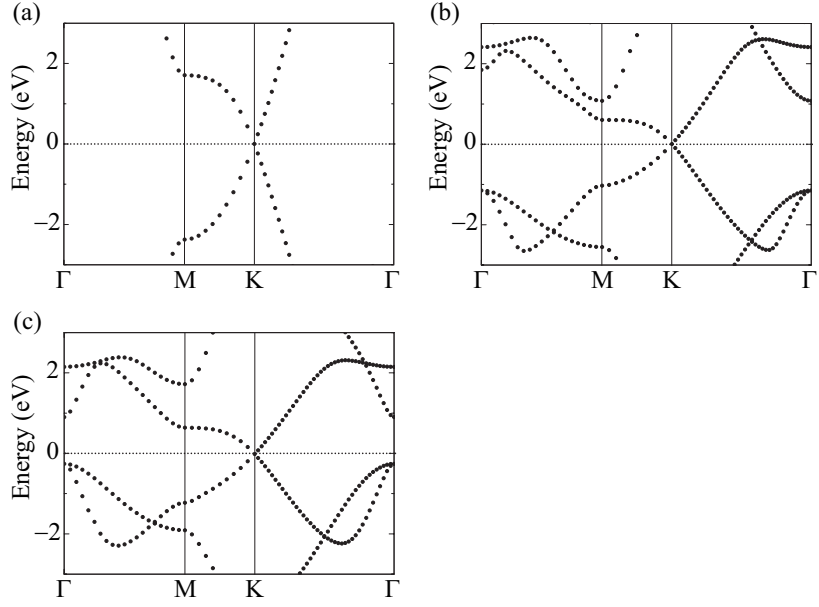}
\end{center}
\caption{Band structures for {\bf(a)} graphene, {\bf(b)} silicene, and {\bf(c)} germanene. The zero of energy is chosen to be the Fermi level.}
\label{fig:ono-band}
\end{figure}

\newpage
\begin{figure}[tb]
\begin{center}
\includegraphics[width=12cm]{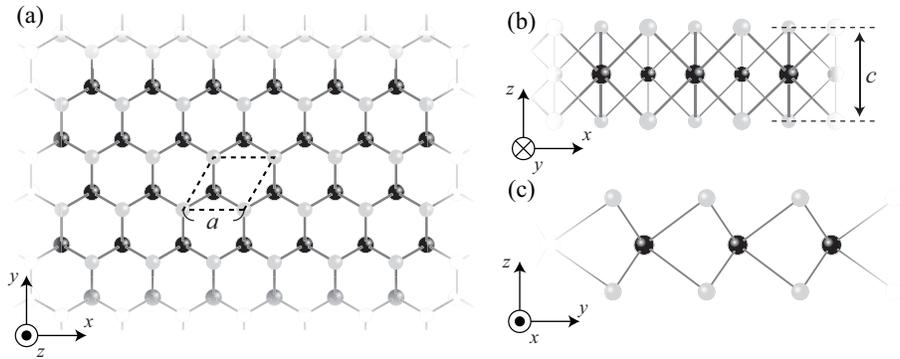}
\end{center}
\caption{Schematic view of 2D MoS$_2$ monolayer. Black and light gray spheres represent Mo and S atoms, respectively. {\bf(a)} is the top view and the parallelogram drawn with broken lines represents a unit cell. {\bf(b)} and {\bf(c)} are the side views of the monolayer.}
\label{fig:egami01}
\end{figure}

\newpage
\begin{figure}[p]
\begin{center}
\includegraphics[width=6cm]{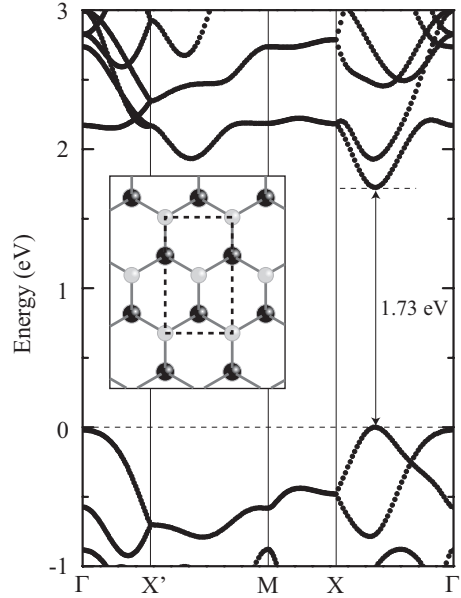}
\end{center}
\caption{Electronic band structure of 2D MoS$_2$ monolayer. The top of the VB is set to be zero. In the inset, the computational model is shown and the rectangular supercell is drawn by broken lines. The key to the symbols in the inset is the same as in Fig.~\ref{fig:egami01}.}
\label{fig:egami04}
\end{figure}

\newpage
\begin{figure}[p]
\begin{center}
\includegraphics[width=12cm]{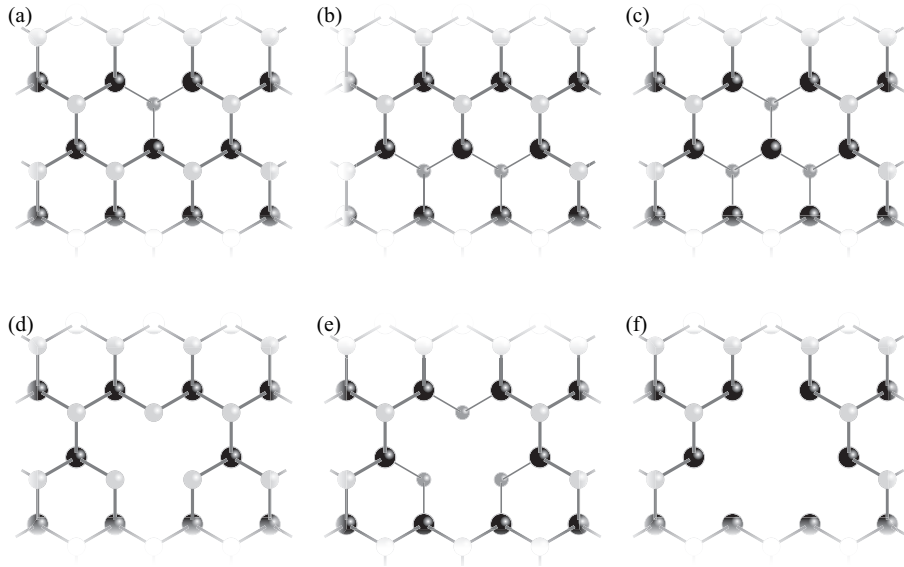}
\end{center}
\caption{Schematic view of vacancy-defect introduced 2D MoS$_2$ monolayer. {\bf(a)} V$_{\rm S}$, {\bf(b)} V$_{\rm S_2}$, {\bf(c)} V$_{\rm S_3}$, {\bf(d)} V$_{\rm Mo}$, {\bf(e)} V$_{\rm MoS_3}$, and {\bf(f)} V$_{\rm MoS_6}$. The key to the symbols is the same as in Fig.~\ref{fig:egami01} and dark gray spheres represent bottom-side S atoms.}
\label{fig:egami02}
\end{figure}

\newpage
\begin{figure}[p]
\begin{center}
\includegraphics[width=6cm]{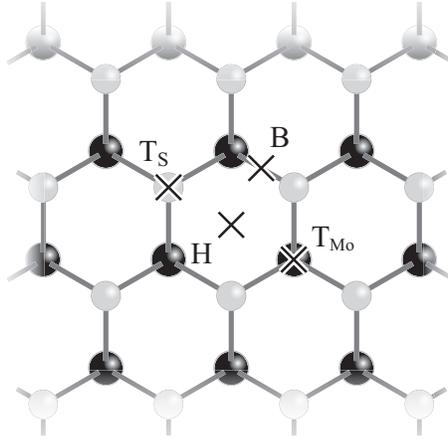}
\end{center}
\caption{Schematic view of adsorption sites on surface of 2D MoS$_2$ monolayer. The sites are indicated by T$_{\mathrm{S}}$ (an on-top site above a S atom), T$_{\mathrm{Mo}}$ (an on-top site above a Mo atom), H (a hollow site), and B (a bridge site). The key to the symbols is the same as in Fig.~\ref{fig:egami01}.}
\label{fig:egami03}
\end{figure}


\newpage
\bibliography{manuscript}

\end{document}